\documentclass[12pt,preprint]{aastex}

\slugcomment{Draft version October 4, 2005.}

\shorttitle{UV spectra of Extreme Helium Stars}
\shortauthors{Pandey et al.}

\begin{document}

\title{An analysis of ultraviolet spectra of Extreme Helium 
Stars and new clues to their origins\altaffilmark{1} 
}

\author{Gajendra Pandey}
\affil{Indian Institute of Astrophysics;
Bangalore,  560034 India}
\email{pandey@iiap.res.in}

\author{David L.\ Lambert}
\affil{Department of Astronomy; University of
Texas; Austin, TX 78712-1083}
\email{dll@astro.as.utexas.edu}

\author{C. Simon Jeffery}
\affil{Armagh Observatory; College Hill,
Armagh BT61 9DG, UK}
\email{csj@arm.ac.uk}

\and

\author{N. Kameswara Rao}
\affil{Indian Institute of Astrophysics;
Bangalore,  560034 India}
\email{nkrao@iiap.res.in}


\altaffiltext{1}{Based on observations obtained with the NASA/ESA
{\it Hubble Space Telescope}, which is operated by the Association of
Universities for Research in Astronomy, Inc. (AURA) under NASA contract
NAS 5-26555} 

\begin{abstract}

Abundances of about 18 elements including the heavy elements
Y and Zr  are determined from  $Hubble$ $Space$ $Telescope$ 
Space Telescope Imaging Spectrograph
ultraviolet spectra of seven extreme helium stars (EHes): LSE\,78,
BD\,+10$^{\circ}$\,2179, V1920\,Cyg, HD\,124448, PV\,Tel,
LS\,IV-1$^{\circ}$\,2, and FQ\,Aqr. 
New optical spectra of the three stars --
BD\,+10$^{\circ}$\,2179, V1920\,Cyg, and HD\,124448 -- were
analysed, and published line lists of 
LSE\,78, HD\,124448, and PV\,Tel were analysed afresh. 
The  abundance analyses is done using LTE line formation
and LTE model atmospheres especially constructed for these EHe stars.
The stellar parameters derived from an EHe's
UV spectrum are in satisfactory agreement with those derived from its
optical spectrum. Adopted abundances for the seven EHes are from
a combination of the UV and optical analyses. 
Published  results  for an additional ten EHes
provide  abundances obtained in a nearly uniform manner for
a total of 17 EHes, the largest sample on record.

The initial metallicity of an EHe is indicated by the abundance of elements from
Al to Ni; Fe is adopted to be the representative of initial metallicity. 
Iron abundances range from
approximately solar to about one-hundredth of solar. Clues to EHe evolution
are contained within the H, He, C, N, O, Y, and Zr abundances. 
Two novel results are (i) the O abundance for some stars is close to the
predicted initial abundance yet the N abundance indicates almost
complete conversion of initial C, N, and O to N by the CNO-cycles;
(ii) three of the seven stars with UV spectra show a strong enhancement
of Y and Zr attributable to an $s$-process.

The observed compositions are discussed in light of expectations from
accretion of a He white dwarf by a CO white dwarf. Qualitative
agreement seems likely except that a problem may be presented by those
stars in which the O abundance is close to the initial O abundance.

\end{abstract}

\keywords{stars: abundances --
stars: chemically peculiar -- stars: evolution}

\section{Introduction}

The extreme helium stars whose chemical compositions are the subject
of this paper are a rare class of peculiar stars.
There are about 21 known EHes. They are supergiants with effective
temperatures in the range 9000 -- 35,000 K and in which surface
hydrogen is effectively a trace element, being underabundant by a factor
of 10,000 or more. Helium is the most abundant
element. Carbon is often the second most abundant element with 
C/He $\simeq$ 0.01,  by number. Nitrogen is overabundant with
respect to that expected for the EHe's metallicity. Oxygen abundance varies from
star to star
but C/O $\simeq$ 1 by number is the maximum ratio found in some
examples. Abundance analyses of varying degrees of completeness
have been reported for a majority of the known EHes. 
The chemical composition should be a primary
constraint on theoretical interpretations of the origin and
evolution of EHes. 

Abundance analyses were first reported by \citet{hill65} for three EHes
by a curve-of-growth technique. Model atmosphere based analyses
of the same three  EHes
were subsequently  reported by
\citet{sch74}, \citet{heb83} and  \citet{sch86}.
\citet{jeff96} summarized the available results for
about  11 EHes. More recent work includes that by
\citet{har97}, \citet{jeff97}, \citet{drill98}, \citet{jeff98},
\citet{jef98}, \citet{jeff99}, and
\citet{pan01}. \citet{rao05a} reviews the results available
for all these stars.

In broad terms, the chemical compositions suggest a hydrogen
deficient atmosphere now composed of material exposed to both 
H-burning and He-burning.
However, the coincidence of  H-processed and He-processed material
at the stellar surface presented a puzzle for many years. Following
the elimination of several proposals, two principal theories emerged:
the `double-degenerate' (DD) model and the `final-flash' (FF) model.

The `double-degenerate' (DD) model was  proposed by \citet{webb84} and \citet{iben84}
and involves merger of a He white dwarf with a more massive C-O white dwarf 
following the decay of their orbit. The binary began life as a close 
pair of normal main sequence stars which through two episodes of mass transfer 
evolved to a He and C-O white dwarf. Emission of gravitational radiation leads 
to orbital decay and to a  merger of the less massive helium
white dwarf with its companion. As a result of the merger the
helium white dwarf is destroyed and forms a thick disk around the more
massive C-O companion. The merging process lasting a few minutes is completed  
as the thick disk is accreted by the C-O white dwarf.
If the mass of the former C-O white dwarf remains below the Chandrasekhar
limit, accretion ignites the base of the accreted envelope forcing the envelope to
expand to supergiant dimensions. Subsequently, it will appear probably first as a cool
hydrogen-deficient carbon star (HdC) or a R Coronae Borealis star (RCB).
 As this H-deficient supergiant contracts, it will become an EHe before
cooling to become a single white dwarf.
(If the merger increases the C-O white dwarf's mass over the Chandrasekhar
limit, explosion as a SN Ia or formation of a neutron star occurs.)

Originally described in quite general terms
\citep{webb84,iben84}, detailed evolution models were
computed only recently (Saio \& Jeffery 2002). The latter included
predictions of the surface abundances of hydrogen, helium, carbon,
nitrogen and oxygen of the resultant EHe.
 A comparison between predictions of the DD model and
observations of EHe's with respect to luminosity to mass ratios ($L/M$),
evolutionary contraction rates,
pulsation masses, surface abundances of H, C, N, and O, and the number
of EHes in the Galaxy concluded that the DD model was the preferred origin for the
EHes and, probably, for the majority of RCBs.
The chemical similarity and the commonality of $L/M$ ratios had long
suggested an evolutionary connection between the EHes and the RCBs
\citep{sch77,rao05a}.

Saio \& Jeffery's (2002) models do not consider the
chemical structure of the white
dwarfs and the EHe  beyond the principal elements (H, He, C, N and O), nor
do they compute the full hydrodynamics of the merger
process and any attendant nucleosynthesis.
Hydrodynamic simulations have been addressed by
{\it inter alia} \citet{hac86}, \citet{benz90},
\citet{seg97}, and \citet{gue04}. Few of the considered
cases involved a He and a C-O white dwarf.  In one example
described by Guerrero et al., a 0.4$M_\odot$ He white dwarf
merged with a 0.6$M_\odot$ C-O white dwarf with negligible mass loss
over the 10 minutes required for complete acquisition of the He white
dwarf by the C-O white dwarf. Accreted material was heated sufficiently
that nuclear burning occurs, mostly by $^{12}$C$(\alpha,\gamma)^{16}$O,
but is quickly quenched. It would appear that negligible nucleosynthesis
occurs in the few minutes  that elapse during the merging.

The second model, the FF model, refers to a late or final
He-shell flash in a post-AGB star which may account for
some  EHes and RCBs. 
In this model \citep{iben83}, the ignition of the  helium shell
in a post-AGB star, say, a cooling white dwarf, results in
what is known as  a late or very late thermal
pulse \citep{her01}. The outer layers  expand rapidly to giant
dimensions. If the hydrogen in the envelope is consumed by H-burning, the
giant becomes a H-deficient supergiant and then contracts  to become an EHe.
The FF model accounts well for several unusual objects
including, for example, FG\,Sge \citep{herb68,lang74,gonz98} 
and V4334\,Sgr (Sakurai's object) \citep{duer96,asp97b}, hot Wolf-Rayet central
stars, and the very hot PG1159 stars \citep{wer91,leu96}.


Determination of surface compositions of EHes
should be rendered as complete as possible: many elements and many stars.
Here, a step is taken toward a more complete specification of the
composition of seven EHes. The primary motivation of our project
was to establish the abundances of key elements heavier than
iron in order to measure the $s$-process concentrations. These
elements are unobservable in the optical spectrum of a hot EHe but
tests showed a few elements should be detectable in ultraviolet spectra.
A successful pilot study of two EHes with the prime motive to measure
specifically the abundances of key elements heavier than iron was reported 
earlier \citep{pan04}.
We now extend the study to all seven stars and to all the elements with
useful absorption lines in the observed UV spectral regions. 
In the following sections, we describe the ultraviolet and optical spectra,
the model atmospheres and the abundance analysis, and discuss the derived chemical
compositions  in light of the DD model.


\section{Observations}

A primary selection criterion for inclusion of an EHe in our program was
its UV flux because useful lines of the heavy elements lie in the UV.
Seven EHes were  observed with the {\it Hubble Space Telescope} and the
{\it Space Telescope Imaging Spectrometer} ({\it STIS}).
  The log of the observations
is provided in Table 1. Spectra were acquired with {\it STIS} using the
E230M grating and the $0.^"2 \times 0.^"06$ aperture. The spectra cover the
range from 1840 \AA\ to 2670 \AA\ at a resolving power 
($R = \lambda/\Delta\lambda$) of 30,000. The raw recorded spectra were 
reduced using the standard {\it STIS} pipeline.  A final spectrum for each
EHe was obtained by co-addition of two or three individual spectra.
Spectra of each EHe in the intervals 2654 \AA\ to
2671 \AA\ and 2401 \AA\ to 2417 \AA\ illustrate the quality and diversity 
of the spectra (Figures 1 and 2), principally the increasing strength and
number of absorption lines with decreasing effective temperature.

New optical
spectra of BD\,+10$^{\circ}$\,2179, and V1920 Cyg were acquired
 with the W.J. McDonald
Observatory's 2.7-m Harlan J. Smith telescope and the coud\'{e} cross-dispersed
echelle spectrograph \citep{tull95} at resolving powers of 45,000 
to 60,000. The observing procedure and wavelength coverage were described
by \citet{pan01}.

Finally, a spectrum of HD\,124448 was obtained with the Vainu Bappu Telescope
of the Indian Institute of Astrophysics with a fiber-fed cross-dispersed
echelle spectrograph \citep{rao04,rao05b}.
The  1000\AA\ of spectrum in 50\AA\ intervals of 30 echelle orders
from 5200 \AA\ to nearly 10,000 \AA\ was recorded on a Pixellant
CCD. The resolving power was about 30,000. The S/N in the continuum
was 50 to 60.

\clearpage
\begin{figure}
\epsscale{1.00}
\plotone{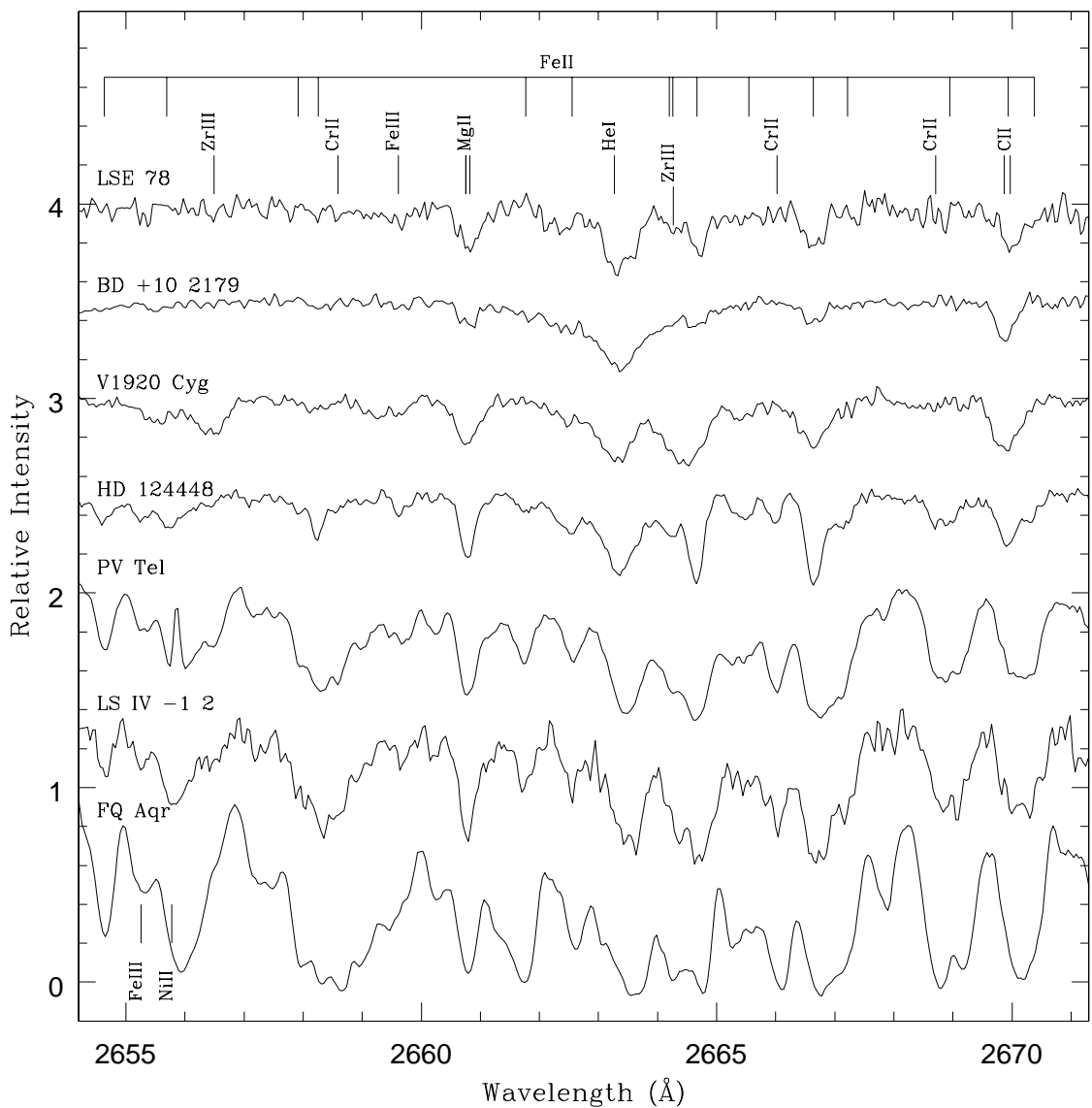}
\caption{A sample of the {\it STIS} spectra of the seven EHes.
The spectra are normalized to the continuum and are shown with offsets 
of about 0.5 between each. Several
lines are identified in this window from 2654 \AA\ to 2671 \AA. Stars
are arranged from top to bottom in order of decreasing effective
temperature. \label{fig1}}
\end{figure}

\begin{figure}
\epsscale{1.00}
\plotone{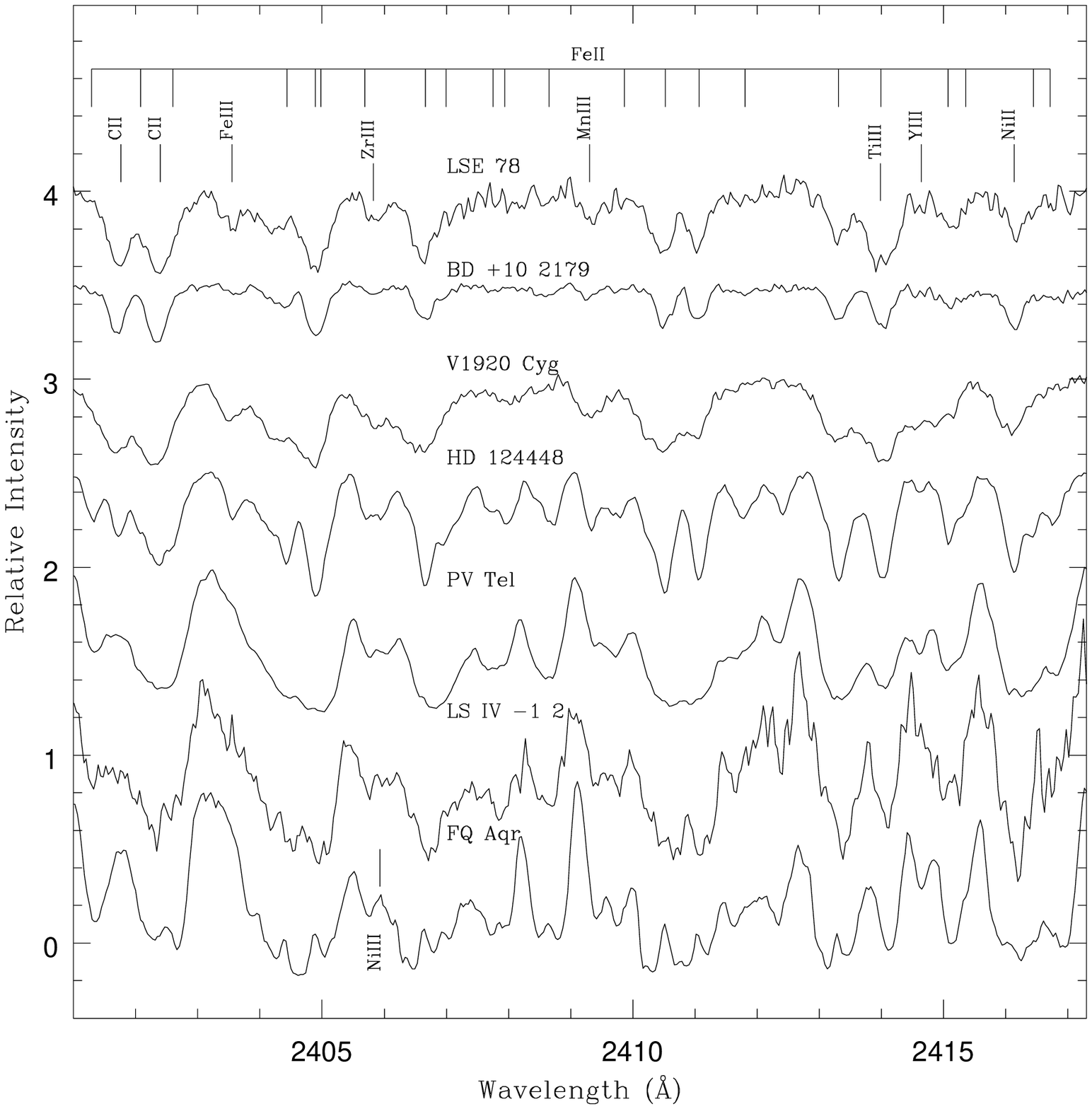}
\caption{A sample of the {\it STIS} spectra of the seven EHes. 
The spectra are normalized to the continuum and are shown with offsets 
of about 0.5 between each. Several
lines are identified in this window from 2401 \AA\ to 2417 \AA. Stars
are arranged from top to bottom in order of decreasing effective
temperature. \label{fig2}}
\end{figure}
\clearpage
\begin{deluxetable}{lccccc}
\tabletypesize{\scriptsize}
\tablewidth{0pt}
\tablecolumns{6}
\tablecaption{The $HST$ $STIS$ Observations}
\tablehead{
\colhead{Star}      & \colhead{$V$} & \colhead{Obs. Date} &
\colhead{Exp. time} & \colhead{S/N} & \colhead{Data Set Name} \\
\colhead{} & \colhead{mag} & \colhead{} & \colhead{s} & \colhead{at 2500\AA} &
\colhead{} }
\startdata
V2244\,Oph & 11.0 &         &         & 28 & \\
($=$LS\,IV-1$^{\circ}$\,2)& & 7 Sep 2002 & 1742 &    & O6MB04010\\
                          & & 7 Sep 2002 & 5798 &    & O6MB04020\\
&&&&&\\
BD+1$^{\circ}$\,4381 & 9.6 &    &     & 59 & \\
($=$FQ\,Aqr)& & 10 Sep 2002 & 1822 & & O6MB07010 \\
            & & 10 Sep 2002 & 5798 & & O6MB07020 \\
&&&&&\\
HD\,225642 & 10.3 &         &         & 45 & \\
($=$V1920\,Cyg)& & 18 Oct 2002 & 1844 & & O6MB06010 \\
               & & 18 Oct 2002 & 2945 & & O6MB06020 \\
&&&&&\\
BD\,+10$^{\circ}$\,2179 & 10.0 &      &     & 90 & \\
               & & 14 Jan 2003 & 1822 & & O6MB01010 \\
               & & 14 Jan 2003 & 2899 & & O6MB01020 \\
&&&&&\\
CoD\,-46$^{\circ}$\,11775 & 11.2 &       &    &  50 & \\
($=$LSE\,78)   & & 21 Mar 2003 & 2269 & & O6MB03010 \\
               & & 21 Mar 2003 & 2269 & & O6MB03020 \\
&&&&&\\
HD\,168476 &  9.3 &         &         & 90 & \\
($=$PV\,Tel)  & & 16 Jul 2003 & 2058 & & O6MB05010 \\
              & & 16 Jul 2003 & 3135 & & O6MB05020 \\
&&&&&\\
HD\,124448 & 10.0 &         &         & 70 &  \\
              & & 21 Jul 2003 & 1977 & & O6MB05010 \\
              & & 21 Jul 2003 & 3054 & & O6MB05020 \\
\enddata
\end{deluxetable}
\clearpage

\section{Abundance Analysis -- Method}

\subsection{Outline of the procedure}

The abundance analysis follows closely a procedure
described by \citet{pan01,pan04}.
H-deficient model atmospheres have been computed
using the code STERNE \citep{jeff01} for
the six stars with an effective temperature greater than 10,000 K.
For FQ\,Aqr with $T_{\rm eff} = 8750$ K, we adopt the Uppsala 
model atmospheres \citep{asp97a}. Both
codes include  line blanketing. Descriptions of the line
blanketing and the sources of continuous opacity are given in the
above references. \citet{pan01} showed that the two codes
gave consistent abundances at 9000 --  9500 K, the upper temperature bound for
the Uppsala models and the lower temperature bound for STERNE
models.  Local thermodynamic equilibrium (LTE) is adopted for
all aspects of model construction.

 A model atmosphere is used with the Armagh LTE code SPECTRUM
\citep{jeff01} to compute the equivalent width of a line or
a synthetic spectrum for a selected spectral window. In matching
a synthetic spectrum to an observed spectrum we include broadening
due to the instrumental profile, the microturbulent velocity $\xi$ and assign all
additional broadening, if any, to rotational broadening.
In the latter case, we use the standard rotational broadening
function $V(v\sin i,\beta)$ \citep{uns55,duft72} with
the limb darkening coefficient set at $\beta = 1.5$.
Observed unblended line profiles are used to obtain the projected rotational 
velocity $v\sin i$. We find that the synthetic line profile, including
the broadening due to instrumental profile, for
the adopted model atmosphere ($T_{\rm eff}$,$\log g, \xi$) and the abundance 
is sharper than the observed. This extra broadening in the observed profile
is attributed to rotational broadening. 
Since we assume that macroturbulence is vanishingly small, 
the $v\sin i$ value is an upper limit to the true value.

\setcounter{footnote}{1}
The adopted $gf$-values are from
the NIST database\footnote{http://physics.nist.gov/cgi-bin/AtData/}, 
\citet{wies96}, \citet{ekb97},
\citet{uyl97}, \citet{raas97}, \citet{mart88},
\citet{art81}, \citet{cresp94}, \citet{sali85},
Kurucz's database\footnote{http://kurucz.harvard.edu}, 
and the compilations by R. E. Luck (private communication).
The adopted $gf$-values for Y\,{\sc iii}, Zr\,{\sc iii}, La\,{\sc iii},
Ce\,{\sc iii}, and Nd\,{\sc iii}, are discussed in \citet{pan04}.
The Stark broadening and radiative broadening coefficients, if available, are mostly
taken from the Vienna Atomic Line 
Database\footnote{http://www.astro.univie.ac.at/$\sim$vald}.
The data  for
computing He\,{\sc i} profiles are the same as in \citet{jeff01},
except for the He\,{\sc i} line at 6678\AA, for which 
the $gf$-values and electron broadening coefficients 
are from Kurucz's database. The line broadening coefficients
are not available for the He\,{\sc i} line at 2652.8\AA.
Detailed line lists used in our analyses are available in electronic form.

\subsection{Atmospheric parameters}

The model atmospheres are characterized by the effective temperature, the
surface gravity, and the chemical composition. 
A complete iteration on chemical composition was not undertaken, i.e.,
the input composition was not fully consistent with the composition
derived from the spectrum with that model. Iteration was done
for the He and C abundances which, most especially He, dominate the
continuous opacity at optical and UV wavelengths. Iteration was
not done for the elements (e.g., Fe -- see Figures 1 and 2) which
contribute to the line blanketing.

The stellar parameters are determined from the line spectrum.
The microturbulent velocity $\xi$ (in km s$^{-1}$) is first determined
by the usual requirement that the abundance from
a set of lines of the same ion be independent of a line's equivalent
width. The result will be insensitive to the assumed effective
temperature provided that the lines span only a small range
in excitation potential.
 For an element represented in the
spectrum by two or more ions, imposition of ionization
equilibrium (i.e., the same abundance is required from lines of
different stages of ionization) defines a locus in the
($T_{\rm eff},\log g)$ plane.
Except for the coolest star in our sample (FQ\,Aqr),
a locus is insensitive to
the input C/He ratio of the model.
 Different pairs of ions of a common element provide
loci of very similar slope in the ($T_{\rm eff},\log g)$ plane.

An indicator yielding a locus with a contrasting slope 
in the ($T_{\rm eff},\log g)$ plane is required to break the 
degeneracy presented by ionization equilibria. 
A potential indicator is a He\,{\sc i} line.
For stars hotter than about 10,000 K,
the  He\,{\sc i} lines are less sensitive to $T_{\rm eff}$
than to $\log g$ on account of pressure broadening due to 
the quadratic Stark effect.
The diffuse series lines are, in particular, useful because they are
less sensitive to the microturbulent velocity than the sharp lines.
A second indicator may be available: species represented
by lines spanning a range in excitation potential may serve as
a thermometer measuring $T_{\rm eff}$ with a weak dependence
on $\log g$. 
 
For  each of the  seven stars, a published abundance analysis 
gave estimates of the atmospheric parameters.
We took these estimates as initial values for the analysis
of our spectra.

\section{Abundance Analysis -- Results}

The seven stars are discussed one by one from hottest to
coolest. Inspection of Figures 1 and 2 shows that many lines
are resolved and only slightly blended in the hottest four
stars. The coolest three stars are rich in lines and
spectrum synthesis is a necessity in determining the
abundances of many elements. 

The hotter stars of our sample have a well defined continuum, the region
of the spectrum (having maximum flux) free of absorption lines is treated
as the continuum point and a smooth curve passing through these
points (free of absorption lines) is defined as the continuum.
For the relatively less hot stars of our sample, same procedure as above
is applied to place the continuum; for the regions which are severely crowded
by absorption lines, the continuum of the hot stars is used as a
guide to place the continuum in these crowded regions of the spectra.
These continuum normalised observed spectra are
also compared with the synthetic spectra to judge the continuum of
severely crowded regions.
However, extremely crowded regions for e.g., of FQ\,Aqr are not used
for abundance analysis. 

Our ultraviolet analysis is mainly by
spectrum synthesis, but, we do measure equivalent widths of unblended
lines to get hold of the microturbulent velocity.
However, the individual lines from an ion which contribute significantly
to the line's equivalent width ($W_{\lambda}$) are synthesized including
the adopted mean abundances of the minor blending lines. The abundances 
derived, including the predicted $W_{\lambda}$ for these derived
abundances, for the best overall fit to the observed line profile
are in the detailed line list, except for most of the optical lines 
which have the measured equivalent widths.
Discussion of the UV spectrum is followed by comparisons with the 
abundances derived from the optical spectrum and the presentation
of adopted set of abundances.
Detailed line lists (see for SAMPLE Table 2 which lists some lines
of BD\,+10$^{\circ}$\,2179) used in our analyses lists
the line's lower excitation potential ($\chi$),
$gf$-value, log of Stark damping constant/electron number density ($\Gamma_{el}$),
log of radiative damping constant ($\Gamma_{rad}$), and the abundance derived 
from each line for the adopted model atmosphere. 
Also listed are the equivalent widths ($W_{\lambda}$)
corresponding to the abundances derived by spectrum synthesis for most individual lines.
The derived stellar parameters of the
adopted model atmosphere are accurate to typically: 
$\Delta$$T_{\rm eff}$ = $\pm$500 K, $\Delta$$\log g$ = $\pm$0.25 cgs 
and $\Delta$$\xi$ = $\pm$1 km $s^{-1}$. The abundance error
due to the uncertainty in $T_{\rm eff}$ is estimated by taking a
difference in abundances derived from the 
adopted model ($T_{\rm eff}$,$\log g, \xi$) and a
model ($T_{\rm eff}$$+$500 K,$\log g, \xi$). Similarly, the abundance
error due to the uncertainty in $\log g$ is estimated by taking a
difference in abundances derived from the 
adopted model ($T_{\rm eff}$,$\log g, \xi$) and a
model ($T_{\rm eff}$,$\log g + 0.25, \xi$). The rms error in the
derived abundances from each species for our sample due to the uncertainty 
in $T_{\rm eff}$ and $\log g$ of the derived stellar parameters are in
the detailed line lists.
The abundance errors due to the uncertainty in $\xi$ are not significant,
except for some cases where the abundance is based on one or a few strong lines
and no weak lines, when compared to that 
due to uncertainties in $T_{\rm eff}$ and $\log g$.
These detailed line lists are available in electronic form and also include the 
mean abundance, the line-to-line scatter, the first entry (standard deviation due 
to several lines belonging to the same ion), and for comparison,
the rms abundance error (second entry) from the uncertainty in the adopted 
stellar parameters. The Abundances are given as $\log\epsilon$(X) 
and normalized with respect to $\log\Sigma\mu_{\rm X}\epsilon$(X) $=$ 12.15
where $\mu_{\rm X}$ is the atomic weight of element X.
\clearpage
\begin{deluxetable}{lccccccc}
\tabletypesize{\small}
\tablewidth{0pt}
\tablecolumns{8}
\tablewidth{0pc}
\tablecaption{SAMPLE lines for BD\,+10$^{\circ}$\,2179, the complete line lists for the
seven stars are present in the electronic version of the journal (see Appendix A)}
\tablehead{
\multicolumn{1}{l}{Ion} & \colhead{} & \colhead{} & \colhead{} & \colhead{} \\
\multicolumn{1}{l}{$\lambda$(\AA)} & \multicolumn{1}{c}{log$gf$} & \multicolumn{1}{c}{$\chi$(eV)} &
\multicolumn{1}{c}{$\Gamma_{el}$} & \multicolumn{1}{c}{$\Gamma_{rad}$} &
\multicolumn{1}{c}{$W_{\lambda}$(m\AA)} &
\multicolumn{1}{c}{log $\epsilon$} & \multicolumn{1}{c}{Ref$^a$} \\
}
\startdata
H\,{\sc i} & & & &&&&\\
4101.734 &--0.753 & 10.150 &    &8.790 &Synth&     8.2   &Jeffery \\
4340.462 &--0.447 & 10.150 &    &8.790 &Synth&     8.2   &Jeffery \\
4861.323 &--0.020 & 10.150 &    &8.780 &Synth&     8.5   &Jeffery \\
\hline
Mean:    &        &         &          & &&&8.30$\pm$0.17$\pm$0.20\\
\hline
He\,{\sc i} & & & &&&& \\
4009.260 &-1.470 & 21.218  &&   &Synth&    11.54 & Jeffery  \\
5015.680 &-0.818 & 20.609  &--4.109 &  8.351 &Synth&    11.54 & Jeffery \\
5047.740 &-1.588 & 21.211  &--3.830 &  8.833 &Synth&    11.54 & Jeffery \\
C\,{\sc i} & & & &&&&\\
4932.049 &--1.658 &  7.685 &--4.320 &  &13&     9.3   & WFD \\
5052.167 &--1.303 &  7.685 &--4.510 &  &28&     9.3   & WFD \\
\hline
Mean:    &        &         &          & &&&9.30$\pm$0.00$\pm$0.25\\
\hline
C\,{\sc ii} & & & &&&&\\
3918.980 &--0.533 & 16.333 &--5.042 &  8.788 &286&      9.4  & WFD \\
3920.690 &--0.232 & 16.334 &--5.043 &  8.787 &328&      9.4  & WFD \\
4017.272 &--1.031 & 22.899 & &   &43&      9.3  & WFD \\
4021.166 &--1.333 & 22.899 & &   &27&      9.3  & WFD \\
\enddata
\tablenotetext{a}{Sources of $gf$ values.}
\end{deluxetable}
\clearpage

\subsection{LSE\,78}

\subsubsection{The ultraviolet spectrum}

Analysis of the ultraviolet spectrum began with determinations of
$\xi$.
 Adoption of the model
atmosphere with parameters found by \citet{jeff93} 
gave $\xi$ for 
  C\,{\sc ii}, Cr\,{\sc iii},
and Fe\,{\sc iii} (Figure 3): $\xi \simeq 16\pm1$
km s$^{-1}$. Figure 3 illustrates the method for obtaining the
microturbulent velocity in LSE\,78 and other stars.
\begin{figure}
\epsscale{1.00}
\plotone{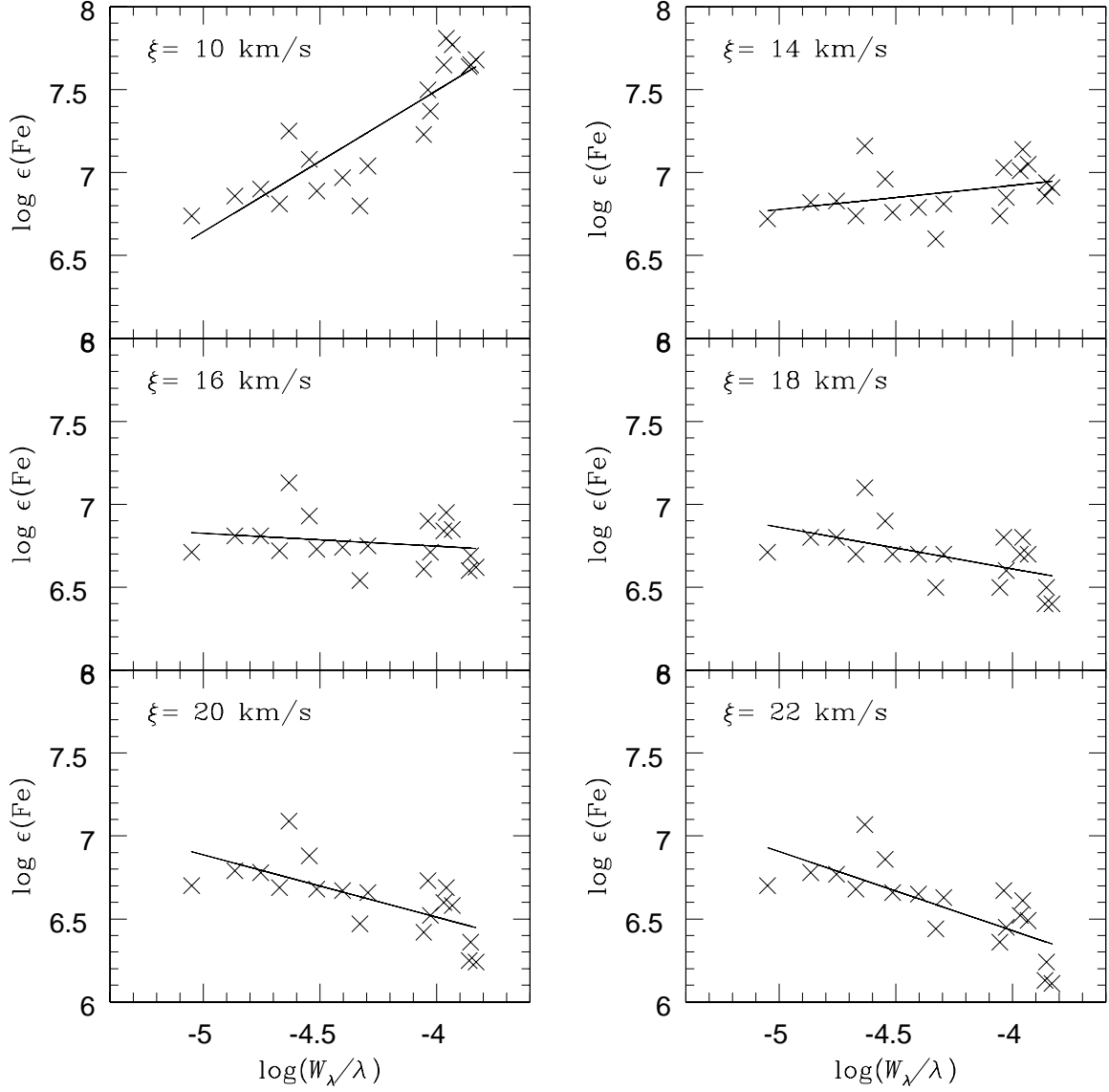}
\caption{Abundances from Fe\,{\sc iii} lines for LSE\,78 versus their
reduced equivalent widths (log $W_{\lambda}/\lambda$).
A microturbulent velocity of $\xi \simeq 16$ km s$^{-1}$ is obtained from this
figure. \label{fig3}}
\end{figure}
Lines of C\,{\sc ii} and C\,{\sc iii} span a large range in
excitation potential. With the adopted $\xi$, models were found which
give the same abundance independent of excitation potential. Assigning
greater weight to C\,{\sc ii} because of the larger number of lines
relative to just the three C\,{\sc iii} lines, we find
$T_{\rm eff} = 18,300\pm 400$ K.  The result is almost independent of the
adopted surface gravity  for C\,{\sc ii} but somewhat dependent on
gravity for the C\,{\sc iii} lines.
Ionization equilibrium loci for C\,{\sc ii}/C\,{\sc iii},
Al\,{\sc ii}/Al\,{\sc iii}, Fe\,{\sc ii}/Fe\,{\sc iii}, and
Ni\,{\sc ii}/Ni\,{\sc iii} are shown in Figure 4.
These with  the estimate $T_{\rm eff} = 18300$ K indicate that 
$\log g = 2.2\pm 0.2$ cgs.
The locus for
Si\,{\sc ii}/Si\,{\sc iii}  is displaced but is discounted because
the Si\,{\sc iii} lines appear contaminated by emission. The He\,{\sc i}
line at 2652.8 \AA\ provides another locus (Figure 4).
\begin{figure}
\epsscale{1.00}
\plotone{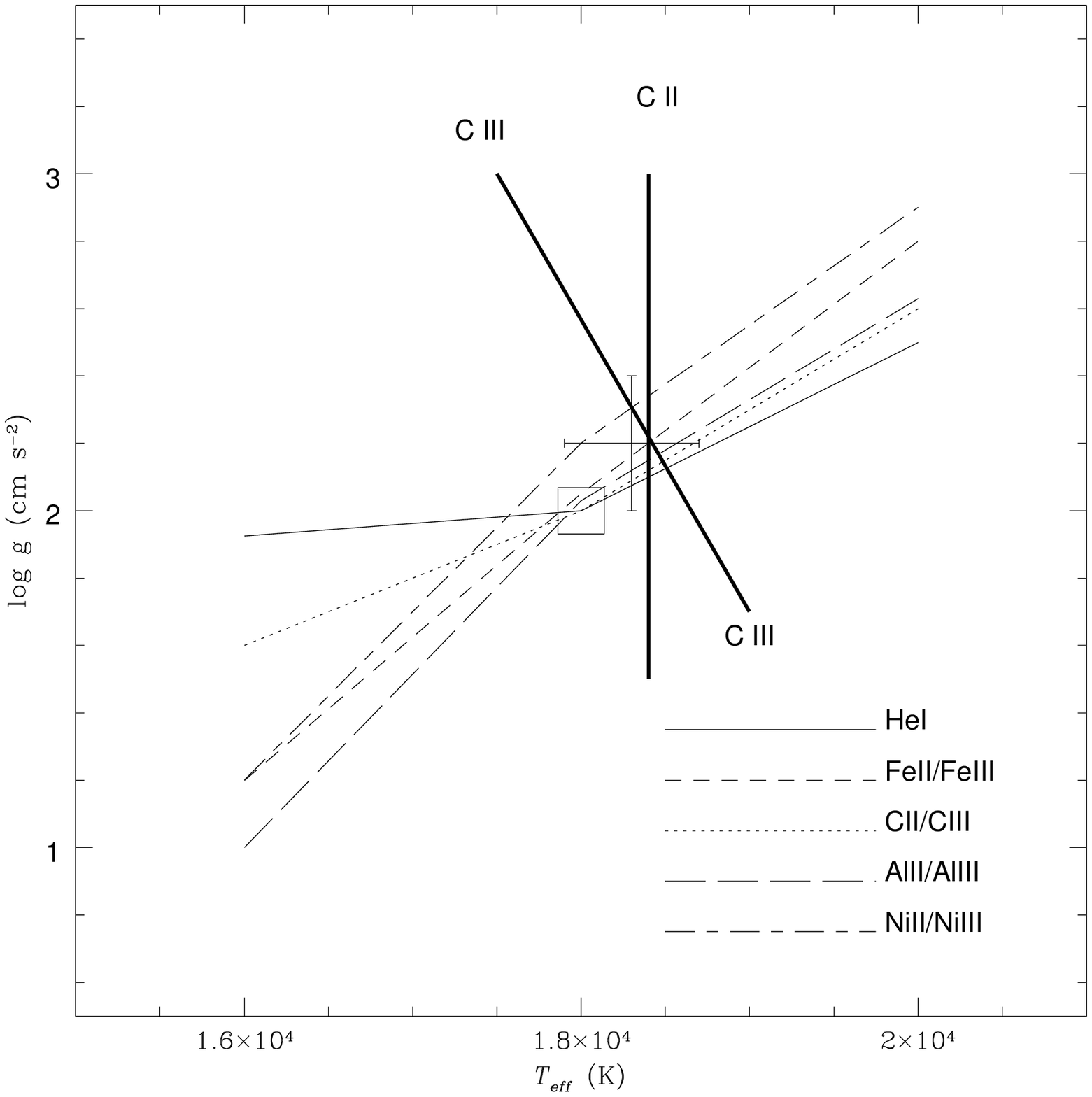}
\caption{The $T_{\rm eff}$ vs $\log g$ plane for LSE\,78. Loci satisfying
ionization equilibria are plotted -- see key on the figure. The locus
satisfying the He\,{\sc i} line profile is shown by the solid
line. The loci satisfying the excitation balance of C\,{\sc ii} 
and C\,{\sc iii} lines are shown by thick solid lines. 
The cross shows the adopted model atmosphere parameters. 
The large square shows the parameters chosen by
\citet{jeff93} from his analysis of an optical spectrum.
\label{fig4}}
\end{figure}
The abundance analysis was undertaken for a STERNE model with
$T_{\rm eff} = 18,300$ K, $\log g = 2.2$ cgs, and $\xi = 16$ km s$^{-1}$.
At this temperature and across the observed wavelength interval,
helium is the leading opacity source and, hence, detailed knowledge of
the composition is not essential to construct an appropriate model.
Results of the abundance analysis are summarized in Table 3.
The deduced $v\sin i$ is about 20 km s$^{-1}$.

\subsubsection{The optical spectrum}

The previous abundance analysis of this EHe was reported by \citet{jeff93}
who analysed a spectrum  covering the interval 3900 \AA\ -- 4800 \AA\
obtained at a resolving power $R \simeq 20,000$ and recorded on
a CCD. The spectrum was analysed with the same family of models and the
line analysis code that we employ. The atmospheric parameters
chosen by Jeffery were 
$T_{\rm eff} = 18000\pm700$ K, $\log g = 2.0\pm0.1$ cgs, and
$\xi = 20$ km s$^{-1}$, and C/He = 0.01$\pm0.005$.
These parameters were
derived exclusively from the  line  spectrum using
ionization equilibria for He\,{\sc i}/He\,{\sc ii},
C\,{\sc ii}/C\,{\sc iii}, S\,{\sc ii}/S\,{\sc iii}, and
Si\,{\sc ii}/Si\,{\sc iii}/Si\,{\sc iv} and the He\,{\sc i}
profiles.

Jeffery noted, as \citet{heb86} had earlier, that the spectrum
contains emission lines, especially of He\,{\sc i} and C\,{\sc ii}.
The emission appears to be weak and is not identified as
affecting the abundance determinations. A possibly more
severe problem is presented by the O\,{\sc ii} lines which
run from weak to saturated and were the exclusive indicator of
the microturbulent velocity. Jeffery was unable to find a value of $\xi$ that
gave an abundance independent of equivalent width. A value greater
than  30 km s$^{-1}$ was indicated but such a value provided predicted
line widths greater than observed values.

Results of our reanalysis of Jeffery's line list for our model
atmosphere are summarized
in Table 3. Our abundances  differ very little from
those given by Jeffery for his slightly different model.
The oxygen and nitrogen abundances are   based on weak lines; strong lines
give a higher abundance, as noted by Jeffery, and it takes $\xi \simeq
35$ km s$^{-1}$ to render the abundances independent of equivalent
width, a very supersonic velocity. One presumes that non-LTE
effects are responsible for this result.

\subsubsection{Adopted Abundances}

The optical and ultraviolet analyses are in good agreement. 
A maximum difference of 0.3 dex occurs for species
represented by one or two lines.
For Al and Si, higher weight is given to the optical lines
because the ultraviolet Al\,{\sc ii}, Al\,{\sc iii}, and
Si\,{\sc iii} lines are partially blended.
The optical and ultraviolet analyses are largely complementary
in that the  ultraviolet provides a good representation of the
iron-group and the optical more coverage of the elements between
oxygen and the iron-group.
Adopted abundances for LSE\,78 are in Table 4; also given are
solar abundances from Table 2 of \citet{lod03} for comparison. 

\clearpage
\begin{table}
\begin{center}\small{Table 3} \\
\small{Chemical Composition of the EHe LSE\,78}\\
\begin{tabular}{lrccccrccccrcl}
\hline
\hline
& & UV $^{\rm a}$ & & &  &  & Optical $^{\rm b}$  
& & &  &  & (Jeffery) $^{\rm c}$ &\\
\cline{2-4}\cline{7-9}\cline{12-14}
Species &  log $\epsilon$ & & $n$\ \ \ & & &   log $\epsilon$ &  & $n$\ \ \
& & &   log $\epsilon$ &  & $n$  \\
\hline
H\,{\sc i} & \nodata & & \nodata & & &  $< 7.5$ & &1 \ \ &&& $< 7.5$ && 1\\
He\,{\sc i} & 11.54 & & 1 \ \ & & &  \nodata & &\nodata \ \ &&& \nodata && \nodata\\
C\,{\sc ii} & 9.4 && 19 \ \ &&& 9.4 && 7 \ \ &&& 9.5 && 10\\
C\,{\sc iii} & 9.6 && 3 \ \  &&& 9.6 && 3 \ \ &&& 9.6 && 6\\
N\,{\sc ii} & 8.0:&& 1 \ \  &&& 8.3 & &12 \ \ &&& 8.4 && 12\\
O\,{\sc ii} & \nodata && \nodata &&& 9.2 && 60\ \ &&& 9.1 && 72\\
Mg\,{\sc ii} & 7.7 &&  2 \ \  &&& 7.4 && 1\ \ &&& 7.2 && 1\\
Al\,{\sc ii} & 6.0 && 1 \ \  &&& \nodata && \nodata  \ \ &&& \nodata && \nodata\\
Al\,{\sc iii} & 6.0 && 1 \ \ &&& 5.8 && 1\ \ &&& 5.8 && 3\\
Si\,{\sc ii} & 7.2 && 2 \ \ &&& 7.0 && 1\ \ &&& 7.1 && 1\\
Si\,{\sc iii} & 6.7 && 2 \ \ &&& 7.2 && 3\ \ &&& 7.1 && 3\\
Si\,{\sc iv} & \nodata & & \nodata & && 7.3 && 1\ \ &&& 7.1 && 1\\
P\,{\sc iii} & \nodata && \nodata &&& 5.3 && 3\ \ &&& 5.3 && 3\\
S\,{\sc ii} & \nodata && \nodata &&& 7.1 && 3\ \ &&& 7.3 && 3\\
S\,{\sc iii} & \nodata && \nodata &&& 6.9 && 2\ \ &&& 6.8 && 2\\
Ar\,{\sc ii} & \nodata && \nodata &&& 6.5 && 4\ \ &&& 6.6 && 4\\
Ca\,{\sc ii} & \nodata && \nodata &&& 6.3 && 2\ \ &&& 6.3 && 2\\
Ti\,{\sc iii} & 4.3 && 8 \ \ &&& \nodata && \nodata\ \ &&& \nodata && \nodata\\
Cr\,{\sc iii} & 4.7 && 44 \ \ &&& \nodata && \nodata\ \ &&& \nodata && \nodata\\
Mn\,{\sc iii} & 4.4 && 6 \ \ &&& \nodata && \nodata \ \ &&& \nodata && \nodata\\
Fe\,{\sc ii} & 6.8 && 37 \ \ &&& \nodata && \nodata\ \ &&& \nodata && \nodata\\
Fe\,{\sc iii} & 6.9 && 38 \ \ &&& 6.7 && 3\ \ &&& 6.8 && 5\\
Co\,{\sc iii} & 4.4 && 2 \ \ &&& \nodata && \nodata\ \ &&& \nodata && \nodata\\
Ni\,{\sc ii}    & 5.6 & & 13 \ \ &&& \nodata && \nodata\ \ &&& \nodata && \nodata\\
Ni\,{\sc iii} & 5.5 && 2 \ \ &&& \nodata && \nodata\ \ &&& \nodata && \nodata\\
Zn\,{\sc ii}    & $< 4.4$ && 1 \ \ &&& \nodata && \nodata\ \ &&& \nodata && \nodata\\
Y\,{\sc iii} & $< 3.2$ && 1 \ \ &&& \nodata && \nodata\ \ &&& \nodata && \nodata\\
Zr\,{\sc iii} & 3.5 && 4 \ \ &&& \nodata && \nodata\ \ &&& \nodata && \nodata\\
La\,{\sc iii} & $< 3.2$ && 1 \ \ &&& \nodata && \nodata\ \ &&& \nodata && \nodata\\
Ce\,{\sc iii} & $< 2.6$ && 1 \ \ &&& \nodata && \nodata\ \ &&& \nodata && \nodata\\
\hline
\end{tabular}
\end{center}
$^{\rm a}$ This paper for the model ($T_{\rm eff}$,$\log g, \xi$) $\equiv$
 (18300, 2.2, 16.0)\\
$^{\rm b}$ Data from \citet{jeff93} and for the model (18300, 2.2, 16.0)\\
$^{\rm c}$ From \citet{jeff93} for his model (18000, 2.0, 20.0)\\
\end{table}


\begin{deluxetable}{lrrrrrrrr}
\tabletypesize{\small}
\tablewidth{0pt}
\tablecolumns{11}
\tablewidth{0pc}
\setcounter{table} {3}
\tablecaption{Adopted Abundances}
\tablehead{
\colhead{X} & \colhead{Solar$^a$} & \colhead{LSE\,78} & 
\colhead{BD\,+10$^{\circ}$\,2179} & \colhead{V1920\,Cyg}
& \colhead{HD\,124448} & \colhead{PV\,Tel} & 
\colhead{LS\,IV-1$^{\circ}$\,2} & \colhead{FQ\,Aqr}\\
}
\startdata
H           & 12.00 & $<$7.5    &8.3        &$<$6.2     &$<$6.3     & 
$<$7.3      &7.1           &6.2 \\
He          & 10.98 & 11.54     &11.54      &11.50      &11.54      
&11.54      &11.54       &11.54 \\
C           & 8.46  &9.5        &9.4        &9.7        &9.2        
&9.3        &9.3        &9.0    \\
N           & 7.90  &8.3        &7.9        &8.5        &8.6        
&8.6        &8.3        &7.2   \\
O           & 8.76  &9.2        &7.5        &9.7        &8.1        & 
8.6       &8.9        &8.9    \\
Mg          & 7.62  &7.6        &7.2        &7.7        &7.6        
&7.8        &6.9        &6.0    \\
Al          & 6.48  &5.8        &5.7        &6.2        &6.5        
&6.2:       & 5.4       &4.7    \\
Si          & 7.61  &7.2        &6.8        &7.7        &7.1        
&7.0        & 5.9       &6.1   \\
P           & 5.54      &5.3        &5.3        &6.0        &5.2        
&6.1        & 5.1       &4.2   \\
S           & 7.26  &7.0        &6.5        &7.2        &6.9        & 
7.2       &6.7         & 6.0 \\
Ar          &6.64      &6.5        &6.1        &6.5        &6.5        & 
\nodata   &\nodata     & \nodata  \\
Ca          & 6.41  &6.3        &5.2        &5.8        &$<$6.0     & 
\nodata   & 5.8       & 4.2   \\
Ti          & 4.93  &4.3        &3.9        &4.5        &4.8        & 
5.2:      & 4.7       &3.2    \\
Cr          & 5.68  &4.7        &4.1        &4.9        &5.2        & 
5.1       & 5.0       &3.6    \\
Mn          & 5.58  &4.4        &4.0        &4.7        &4.9        
&4.9        & \nodata   &3.9    \\
Fe          & 7.54  &6.8        &6.2        &6.8        &7.2        
&7.0        &6.3        &5.4    \\
Co          & 4.98  &4.4        & \nodata   &4.4        &4.6        
&\nodata    & \nodata   &3.0   \\
Ni          & 6.29  &5.6        &5.1        &5.4        &5.6        
&5.7        & 5.1       &4.0    \\
Cu          & 4.27  &\nodata    & \nodata   & \nodata   & \nodata   & 
\nodata   & \nodata   &2.7    \\
Zn          & 4.70  &$<$4.4     &4.4        &4.5        &\nodata    
&\nodata    & \nodata   &3.2    \\
Y           & 2.28  &$<$3.2     &$<$1.4     &3.2        &2.2        
&2.9        & 1.4       &\nodata    \\
Zr          & 2.67  &3.5        &$<$2.6     &3.7        &2.7        
&3.1        & 2.3     &1.0    \\
La          & 1.25  & $<$3.2    & \nodata   &$<$2.2     & \nodata   
&\nodata    & \nodata   &\nodata    \\
Ce          & 1.68  & $<$2.6    &$<$2.0     &$<$2.0     &$<$1.8     
&$<$1.7     & \nodata   &$<$0.3    \\
Nd          & 1.54  & \nodata   &$<$2.0    &$<$1.8     &\nodata
&\nodata    & $<$0.8    &\nodata    \\
\enddata
\tablenotetext{a}{Recommended solar system abundances from Table 2 of 
\citet{lod03}.}
\end{deluxetable}

\clearpage

\subsection{BD$+10^\circ$ 2179}

\subsubsection{The ultraviolet spectrum}

The star was analysed previously by \citet{heb83} from a combination
of ultraviolet spectra obtained with the {\it IUE} satellite and
photographic spectra covering the wavelength interval
3700 \AA\ to 4800 \AA. Heber's model atmosphere parameters were
$T_{\rm eff} = 16800\pm600$ K, $\log g = 2.55\pm0.2$ cgs, $\xi = 7\pm1.5$
km s$^{-1}$, and C/He $=0.01^{+0.003}_{-0.001}$.

In our analysis,
the microturbulent velocity was determined from  Cr\,{\sc iii}, Fe\,{\sc ii},
and Fe\,{\sc iii} lines. The three ions give
a similar result and a mean value $\xi = 4.5\pm1$ km s$^{-1}$.
Two ions provide lines spanning a large range in excitation potential
and are, therefore, possible thermometers. The $T_{\rm eff} =
16850$ K according to 17 C\,{\sc ii} lines and 17250 K from
two C\,{\sc iii} lines. When weighted by the number of lines, the
mean is $T_{\rm eff} = 16900$ K. The major uncertainty probably
arises from the combined use of a line or two from the ion's ground
configuration with lines from highly excited configurations and
our insistence on the assumption of LTE.
Ionization equilibrium 
for C\,{\sc ii}/C\,{\sc iii},
Al\,{\sc ii}/Al\,{\sc iii}, Si\,{\sc ii}/Si\,{\sc iii},
Mn\,{\sc ii}/Mn\,{\sc iii}, Fe\,{\sc ii}/Fe\,{\sc iii}, and
Ni\,{\sc ii}/Ni\,{\sc iii} 
with the above  effective temperature
gives the estimate $\log g = 2.55\pm0.2$ cgs (Figure 5).
Thus, the abundance analysis was conducted for the model with
$T_{\rm eff} = 16900$ K, $\log g = 2.55$ cgs, and
a microturbulent velocity of  $\xi = 4.5$ km
s$^{-1}$. 
The $v\sin i$ is deduced to be about 18 km s$^{-1}$.
Abundances are summarized in Table 5.

\subsubsection{Optical spectrum}

The spectrum acquired at the McDonald Observatory was
analysed by the standard procedure.
The microturbulent velocity provided by the C\,{\sc ii} lines is
$7.5$ km s$^{-1}$ and by the N\,{\sc ii} lines is 6 km s$^{-1}$.
We adopt 6.5 km s$^{-1}$ as a mean value, a value slightly greater than the
mean of 4.5 km s$^{-1}$ from the ultraviolet lines.
Ionization equilibrium of C\,{\sc i}/C\,{\sc ii}, C\,{\sc ii}/C{\sc iii},
Si\,{\sc ii}/Si{\sc iii}, S\,{\sc ii}/S{\sc iii}, and
Fe\,{\sc ii}/Fe\,{\sc iii} provide nearly parallel and overlapping
loci in the $\log g$ vs $T_{\rm eff}$ plane. Fits to the He\,{\sc i}
lines at 4009 \AA, 4026 \AA, and 4471 \AA\ provide a locus
whose intersection (Figure 5) with the other ionization equilibria suggests a solution
$T_{\rm eff} = 16400\pm500$ K and $\log g = 2.35\pm0.2$ cgs. 
The $v\sin i$ is deduced to be about 20$\pm$2 km s$^{-1}$.
The differences in parameters derived from optical and UV spectra
are within the uncertainties of the determinations. This star does not
appear to be a variable \citep{rao80,hill84,gra84}.
Results of the  abundance analysis  are given in Table 5.

\clearpage
\begin{figure}
\epsscale{1.00}
\plotone{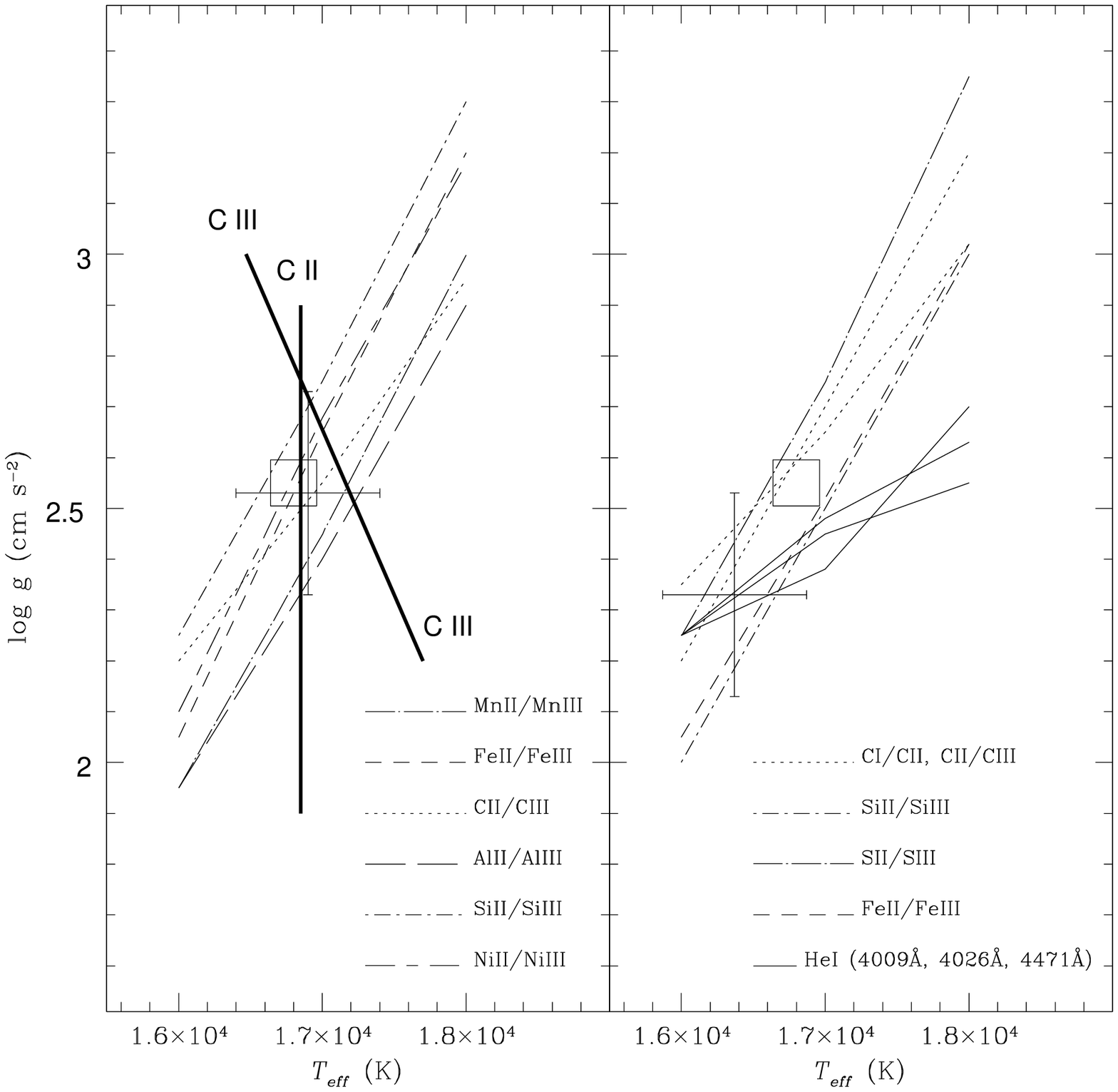}
\caption{The $T_{\rm eff}$ vs $\log g$ plane for BD$+10^\circ 2179$:
the left-hand panel shows the results from the {\it STIS} spectrum,
and the right-hand panel shows results from the optical spectrum. Loci satisfying
ionization equilibria are plotted in both panels -- see keys on the figure.
The loci satisfying  optical He\,{\sc i} line profiles are shown by the solid
lines. The loci satisfying the excitation balance of ultraviolet C\,{\sc ii}
and C\,{\sc iii} lines are shown by thick solid lines in the left-hand panel.
The crosses show the adopted model atmosphere parameters.
The large square shows the parameters chosen by
\citet{heb83}.
\label{fig5}}
\end{figure}
\clearpage

\subsubsection{Adopted abundances}

There is good agreement for common species between the
abundances obtained separately from the ultraviolet and
optical lines. 
Adopted abundances are given in Table 4. These are based on our
$STIS$ and optical spectra.
The N abundance is from the optical N\,{\sc ii} lines because the
ultraviolet N\,{\sc ii}
lines are blended.
The ultraviolet Al\,{\sc iii} line is omitted in forming the
mean abundance because it is very saturated.
The mean Al abundance is gotten from  the  optical Al\,{\sc iii} lines and
the ultraviolet Al\,{\sc ii} lines weighted by the number of lines.
The Si\,{\sc iii} lines are given greater weight than the Si\,{\sc ii}
lines which are generally blended.

Inspection of our abundances showed  large
(0.7 dex) differences for many species between our results and those
reported by \citet{heb83}, see Table 5. We compared Heber's published equivalent widths
and the equivalent widths from our analysis for the common lines.
The differences in equivalent widths found, cannot account for these large
differences in abundances and same is the case for the atomic data ($gf$-values).
This situation led us to
reanalyse Heber's published list of ultraviolet and optical
lines (his equivalent widths and atomic data) using our model
atmosphere. We use the
model $T_{\rm eff} = 16750$ K, and $\log g = 2.5$ cgs,
a model differing by only 50 K and 0.05 cgs from
Heber's choice from a different family of models.
Our estimate of the microturbulent velocity found from  Fe\,{\sc ii}
and Fe\,{\sc iii} lines is about 14 km s$^{-1}$ and
not the 7 km s$^{-1}$ reported by Heber. Heber's value was
obtained primarily from C\,{\sc ii} and N\,{\sc ii}
lines which we found to be unsatisfactory indicators when
using Heber's equivalent widths.
This difference in $\xi$ is confirmed by
a clear trend seen in a plot of equivalent width vs
abundance for Heber's published results for Fe\,{\sc ii}
lines.
This value of $\xi$ is higher than our values from optical
and ultraviolet lines, and higher than the 7 km s$^{-1}$ obtained
by Heber.


Adoption of $\xi  = 14$ km s$^{-1}$,
Heber's equivalent widths and atomic data,
and our model of $T_{\rm eff} = 16750$ K  with
$\log g = 2.5$ provides
abundances very close to our results from the ultraviolet and
optical lines.  Since our optical and UV spectra are of
superior quality to the data available to Heber, we do not
consider our revision of Heber's abundances.
We suspect that
the 14 km s$^{-1}$ for the microturbulent velocity may be an artefact
resulting from a difficulty possibly encountered by Heber in
measuring  weak lines.

\clearpage
\begin{table}
\begin{center}\small{Table 5} \\
\small{Chemical Composition of the EHe BD$+10^\circ$ 2179}\\
\begin{tabular}{lrccccrccccrccccrcc}
\hline
\hline
& & $^{\rm a}$ & & &  &  & $^{\rm b}$ & & &  &  & 
$^{\rm c}$ & & &  &  & $^{\rm d}$ &\\
\cline{2-4}\cline{7-9}\cline{12-14}\cline{17-19}
Species &  log $\epsilon$ & & $n$\ \ \ & & &   log $\epsilon$ &  & $n$\ \ 
\ & & &   log $\epsilon$ &  & $n$\ \ \ & & &   log $\epsilon$ &  & $n$  \\
\hline
H\,{\sc i} & \nodata & & \nodata & & &8.3  & &3 \ \ &&& 8.5 && 2 \ \ &&& 8.6 && 2\\
He\,{\sc i} & \nodata & & \nodata & & &  11.54 & &\nodata\ \ &&& \nodata 
&& \nodata \ \ &&& 11.53 && \nodata\\
C\,{\sc i} & \nodata && \nodata &&& 9.3 && 2 \ \ &&& \nodata && \nodata \ \ &&& \nodata && \nodata\\
C\,{\sc ii} & 9.4 && 29 \ \ &&& 9.3 && 29 \ \ &&& 9.2 && 8 \ \ &&& 9.6 && 22\\
C\,{\sc iii} & 9.5 && 2 \ \  &&& 9.4 && 4 \ \ &&& 9.3 && 1 \ \ &&& 9.6 && 3\\
N\,{\sc ii} & 7.8: && 2 \ \  &&& 7.9 & &28 \ \ &&& 7.7 && 12 \ \ &&& 8.1 && 13\\
O\,{\sc ii} & \nodata && \nodata &&& 7.5 && 11\ \ &&& 7.6 && 4 \ \ &&& 8.1 && 4\\
Mg\,{\sc ii} & 7.2 &&  2 \ \  &&& 7.1 && 2\ \ &&& 7.2: && 2 \ \ &&& 8.0 && 8\\
Al\,{\sc ii} & 5.8 && 2 \ \  &&& \nodata && \nodata  \ \ &&& 5.6 && 2 \ \ &&& 6.3 && 5\\
Al\,{\sc iii} & 6.0 && 1 \ \ &&& 5.6 && 3\ \ &&& 5.4 && 2 \ \ &&& 6.2 && 6\\
Si\,{\sc ii} & 7.0 && 7 \ \ &&& 6.5 && 6\ \ &&& 7.0 && 3 \ \ &&& 7.5 && 4\\
Si\,{\sc iii} & 6.8 && 3 \ \ &&& 6.8 && 5\ \ &&& 6.7 && 5 \ \ &&& 7.3 && 10\\
P\,{\sc ii} & \nodata && \nodata &&& \nodata && \nodata\ \ &&& 5.3 && 3 \ \ &&& 5.4 && 3\\
P\,{\sc iii} & \nodata && \nodata &&& 5.3 && 2\ \ &&& $<$5.1 && 3 \ \ &&& 5.4 && 5\\
S\,{\sc ii} & \nodata && \nodata &&& 6.5 && 15\ \ &&& 7.0 && 8 \ \ &&& 7.2 && 9\\
S\,{\sc iii} & \nodata && \nodata &&& 6.5 && 3\ \ &&& 6.6 && 4 \ \ &&& 7.0 && 4\\
Ar\,{\sc ii} & \nodata && \nodata &&& 6.1 && 3\ \ &&& 6.3 && 3 \ \ &&& 6.4 && 3\\
Ca\,{\sc ii} & \nodata && \nodata &&& 5.2 && 1\ \ &&& 5.4 && 1 \ \ &&& 5.9 && 2\\
Ti\,{\sc iii} & 3.9 && 9 &&& \nodata && \nodata\ \ &&& 3.5 && 8 \ \ &&& 4.1 && 10\\
Cr\,{\sc iii} & 4.1 && 42 &&& \nodata && \nodata\ \ &&& 4.2 && 6 \ \ &&& 5.0 && 8\\
Mn\,{\sc ii} & 4.0 && 4 &&& \nodata && \nodata \ \ &&& $<$4.6 && 3 \ \ &&& $<$4.7 && 3\\
Mn\,{\sc iii} & 4.0 && 27 &&& \nodata && \nodata \ \ &&& 4.1 && 3 \ \ &&& 4.4 && 3\\
Fe\,{\sc ii} & 6.2 && 59 &&& 6.2 && 2\ \ &&& 5.7 && 15 \ \ &&& 6.4 && 16\\
Fe\,{\sc iii} & 6.2 && 67 &&& 6.3 && 6\ \ &&& 5.8 && 22 \ \ &&& 6.5 && 26\\
Co\,{\sc iii} & 4.3: && n &&& \nodata && \nodata\ \ &&& 4.0 && 4 \ \ &&& 4.4 && 4\\
Ni\,{\sc ii}    & 5.1 & & 35 &&& \nodata && \nodata\ \ &&& 5.0 && 2 \ \ &&& 5.2 && 3\\
Ni\,{\sc iii} & 5.1 && 4 &&& \nodata && \nodata\ \ &&& 4.1 && 6 \ \ &&& 5.1 && 6\\
Zn\,{\sc ii}    &4.4      && 1 &&& \nodata && \nodata\ \ &&& \nodata && 
\nodata \ \ &&& \nodata && \nodata\\
Y\,{\sc iii} & $<$1.4 && 2 &&& \nodata && \nodata\ \ &&& \nodata && 
\nodata \ \ &&& \nodata && \nodata\\
Zr\,{\sc iii} & $<$2.6 && 5 &&& \nodata && \nodata\ \ &&& \nodata && 
\nodata \ \ &&& \nodata && \nodata\\
Ce\,{\sc iii} & $< 2.0$ && 1 \ \ &&& \nodata && \nodata\ \ &&& \nodata && 
\nodata \ \ &&& \nodata && \nodata\\
Nd\,{\sc iii} & \nodata && \nodata \ \ &&& $< 2.0$ && 2\ \ &&& \nodata && 
\nodata \ \ &&& \nodata && \nodata\\
\hline
\end{tabular}
\end{center}
$^{\rm a}$ This paper for the model (16900, 2.55, 4.5) from UV lines\\
$^{\rm b}$ This paper for the model (16400, 2.35, 6.5) from optical lines\\
$^{\rm c}$ Rederived from Heber's (1983) list of optical and UV lines
for the model (16750, 2.5, 14.0)\\
$^{\rm d}$ From \citet{heb83} for his model (16800, 2.55, 7.0)\\
\end{table}
\clearpage
\subsection{V1920 Cyg}

An analysis of optical and UV spectra was reported
previously \citep{pan04}. Atmospheric parameters were
taken directly from \citet{jef98} who analysed an
optical spectrum (3900 -- 4800 \AA) using STERNE models and the
spectrum synthesis code adopted here. Here, we report a full
analysis of our {\it STIS} spectrum and the McDonald
spectrum used by Pandey et al.

\subsubsection{The ultraviolet spectrum}

The microturbulent velocity was derived from Cr\,{\sc iii},
Fe\,{\sc ii}, and Fe\,{\sc iii} lines  which gave a value
of 15$\pm$1 km s$^{-1}$. The effective temperature from
C\,{\sc ii} lines was 16300$\pm300K$.
Ionization equilibrium for C\,{\sc ii}/C\,{\sc iii},
Si\,{\sc ii}/Si\,{\sc iii}, Fe\,{\sc ii}/Fe\,{\sc iii}, 
and Ni\,{\sc ii}/Ni\,{\sc iii} provide loci in the
$\log g$ vs $T_{\rm eff}$ plane. The He\,{\sc i} 2652.8 \AA\
profile also provides a locus in this plane.
The final parameters arrived at are (see Figure 6):
$T_{\rm eff} = 16300\pm900$ K, $\log g = 1.7\pm0.35$ cgs,
and $\xi = 15\pm1$ km s$^{-1}$.
 The $v\sin i$ is deduced 
to be about 40 km s$^{-1}$.
The abundances obtained with this model are given in Table 6. 

\subsubsection{The optical spectrum}

The microturbulent velocity from the N\,{\sc ii} lines is
20$\pm1$ km s$^{-1}$. The O\,{\sc ii} lines suggest a
higher microturbulent velocity ($\xi \simeq 24$ km s$^{-1}$ or
even higher when stronger lines are included),
as was the case for LSE\,78. Ionization equilibrium for
S\,{\sc ii}/S\,{\sc iii}, and Fe\,{\sc ii}/Fe\,{\sc iii},
and the fit to He\,{\sc i} profiles for the 4009, 4026, and 4471 \AA\
lines provide loci in the $\log g$ vs $T_{\rm eff}$ plane.
Ionization equilibrium from Si\,{\sc ii}/Si\,{\sc iii} is
not used because the Si\,{\sc ii} lines are affected by
emission.
The final parameters are taken as (see Figure 6):
$T_{\rm eff} = 16330\pm500$ K, $\log g = 1.76\pm0.2$ cgs,
and $\xi = 20\pm1$ km s$^{-1}$.
 The $v\sin i$ is deduced 
to be about 40 km s$^{-1}$. 

The abundance analysis with this model gives the results in Table 6. 
Our abundances  are in fair agreement
with those published by \citet{jef98}. The abundance
differences range from $-$0.5 to $+$0.4 for a mean of 0.1 in the
sense `present study $-$ Jeffery et al.'. A part of the differences may arise from
a slight difference in the adopted model atmospheres. 
\clearpage
\begin{figure}
\epsscale{1.00}
\plotone{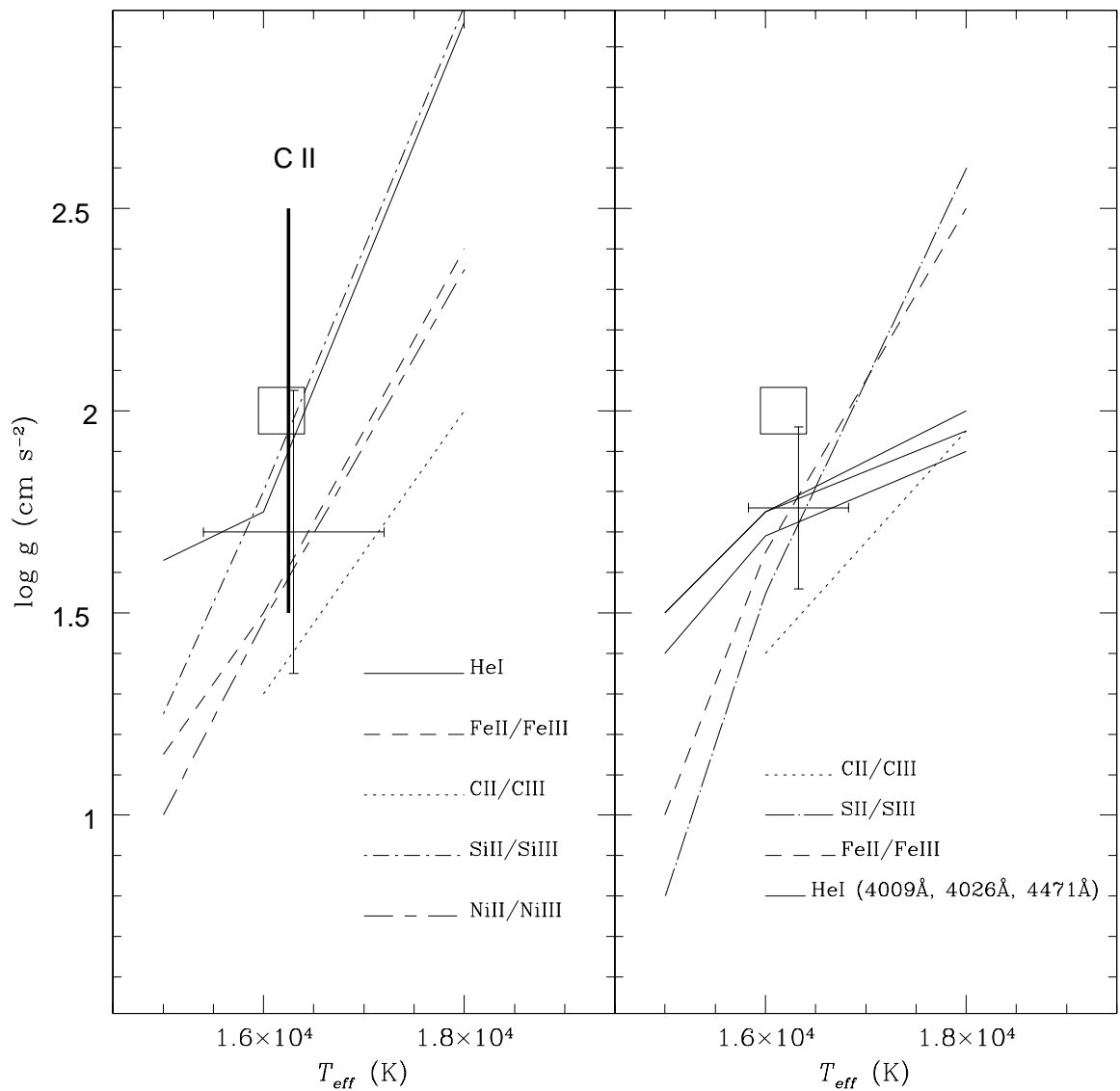}
\caption{The $T_{\rm eff}$ vs $\log g$ plane for  V1920 Cyg:
the left-hand panel shows the results from the {\it STIS} spectrum,
and the right-hand panel shows results from the optical spectrum. Loci satisfying
ionization equilibria are plotted in both panels -- see keys on the figure.
The loci satisfying   He\,{\sc i} line profiles are shown by the solid
lines. The locus satisfying the excitation balance of ultraviolet C\,{\sc ii}
lines is shown by thick solid line in the left-hand panel.
The crosses show the adopted model atmosphere parameters.
The large square shows the parameters chosen by
\citet{jeff98}.
\label{fig6}}
\end{figure}
\clearpage

\subsubsection{Adopted abundances}

Adopted abundances from the combination of $STIS$ and optical
spectra are given in Table 4.
Our limit on the H abundance is from
the absence of the H$\alpha$ line \citep{pan04}.
The C abundance is from ultraviolet and optical C\,{\sc ii} lines
because the ultraviolet  C\,{\sc iii} line is saturated and the
optical C\,{\sc iii} lines are not clean.
The N abundance is from the optical N\,{\sc ii} lines because the
ultraviolet N\,{\sc ii} lines are blended.
The Mg abundance is from the ultraviolet and the optical Mg\,{\sc ii} lines
which are given equal weight.
The ultraviolet Al\,{\sc ii} (blended) and Al\,{\sc iii} (saturated and blended)
lines are given no weight.
The Al abundance is from  the  optical Al\,{\sc iii} lines.
No weight is given to optical Si\,{\sc ii} lines because they are affected
by emissions. The mean Si abundance is from ultraviolet Si\,{\sc ii}, Si\,{\sc iii},
and optical Si\,{\sc iii} lines weighted by the number of lines.
The Fe abundance is from ultraviolet Fe\,{\sc ii}, Fe\,{\sc iii}, and
optical Fe\,{\sc ii}, Fe\,{\sc iii} lines weighted by the number of lines.
Ni abundance is from ultraviolet Ni\,{\sc ii} lines because Ni\,{\sc iii} lines
are to some extent blended. Our adopted abundances are in fair agreement
with Jeffery et al.'s (1998) analysis of their optical spectrum: the mean
difference is 0.2 dex from 11 elements from C to Fe with a difference in model
atmosphere parameters likely accounting for most or all of the differences.
Within the uncertainties, for the common elements, our adopted abundances are also 
in fair agreement with Pandey et al.'s (2004) analysis.

\clearpage
\begin{table}
\begin{center}\small{Table 6} \\
\small{Chemical Composition of the EHe V1920\,Cyg}\\
\begin{tabular}{lrccccrcl}
\hline
\hline
& & UV$^{\rm a}$ & & &  &  & Optical $^{\rm b}$ & \\
\cline{2-4}\cline{7-9}
Species &  log $\epsilon$ & & $n$\ \ \ & & &   log $\epsilon$ &  & $n$  \\
\hline
H\,{\sc i} & \nodata & & \nodata & & &  $< 6.2$ & &1 \\
He\,{\sc i} & 11.54 & & 1 \ \ & & &  11.54 & & 4 \\
C\,{\sc ii} & 9.7 && 11 \ \ &&& 9.6 && 12 \\
C\,{\sc iii} & 9.7 && 1 \ \  &&& 10.4: && 1 \\
N\,{\sc ii} & 8.5:&& 3 \ \  &&& 8.5 & &19 \\
O\,{\sc ii} & \nodata && \nodata &&& 9.7 && 18\\
Mg\,{\sc ii} & 8.0 &&  1 \ \  &&& 7.6 && 2\\
Al\,{\sc ii} & 5.5:&& 1 \ \  &&& \nodata && \nodata  \\
Al\,{\sc iii} & 6.3:&& 1 \ \ &&& 6.2 && 2\\
Si\,{\sc ii} & 7.4 && 2 \ \ &&& 7.0 && 2\\
Si\,{\sc iii} & 7.3 && 1 \ \ &&& 7.9 && 3\\
Si\,{\sc iv} & \nodata & & \nodata & && \nodata && \nodata\\
P\,{\sc iii} & \nodata && \nodata &&& 6.0 && 2\\
S\,{\sc ii} & \nodata && \nodata &&& 7.2 && 10\\
S\,{\sc iii} & \nodata && \nodata &&& 7.3 && 3\\
Ar\,{\sc ii} & \nodata && \nodata &&& 6.5 && 2\\
Ca\,{\sc ii} & \nodata && \nodata &&& 5.8 && 2\\
Ti\,{\sc iii} & 4.5 && 7 \ \ &&& \nodata && \nodata\\
Cr\,{\sc iii} & 4.9 && 41 \ \ &&& \nodata && \nodata\\
Mn\,{\sc iii} & 4.7 && 5 \ \ &&& \nodata && \nodata \\
Fe\,{\sc ii} & 6.7 && 33 \ \ &&& 6.6 && 2 \\
Fe\,{\sc iii} & 6.8 && 25 \ \ &&& 6.8 && 3\\
Co\,{\sc iii} & 4.4 && 2 \ \ &&& \nodata && \nodata\\
Ni\,{\sc ii}    & 5.4 & & 13 \ \ &&& \nodata && \nodata\\
Ni\,{\sc iii} & 5.7: && 2 \ \ &&& \nodata && \nodata\\
Zn\,{\sc ii}    & 4.5 && 1 \ \ &&& \nodata && \nodata\\
Y\,{\sc iii} &  3.2 && 2 \ \ &&& \nodata && \nodata\\
Zr\,{\sc iii} & 3.7 && 6 \ \ &&& \nodata && \nodata\\
La\,{\sc iii} & $< 2.2$ && 1 \ \ &&& \nodata && \nodata\\
Ce\,{\sc iii} & $< 2.0$ && \nodata \ \ &&& \nodata && \nodata\\
Nd\,{\sc iii} & \nodata && \nodata \ \ &&& $< 1.8$ && \nodata\\
\hline
\end{tabular}
\end{center}
$^{\rm a}$ This paper for the model atmosphere (16300, 1.7, 15.0)\\
$^{\rm b}$ This paper for the model atmosphere (16330, 1.8, 20.0)\\
\end{table}
\clearpage

\subsection{HD\,124448}

HD\,124448 was the first EHe star discovered \citep{pop42}.
Membership of the EHe class was opened with Popper's scrutiny of
his spectra of HD\,124448 obtained at the McDonald
Observatory: `no hydrogen lines in absorption or in emission,
although helium lines are strong'. Popper also noted the
absence of a Balmer jump. His attention had been drawn to the
star because faint early-type B stars (spectral type B2 according to the
{\it Henry Draper Catalogue}) are rare at high galactic
latitude. The star is known to cognoscenti  as Popper's star. 

Earlier, we reported an analysis of lines in a limited wavelength
interval of our {\it STIS} spectrum \citep{pan04}. Here, we give a full analysis of
that spectrum.
In addition, we present an analysis of a portion of the optical high-resolution spectrum
obtained with the {\it Vainu Bappu Telescope}.

\subsubsection{The ultraviolet spectrum}

A microturbulent velocity of 10$\pm1$ km s$^{-1}$ is found from Cr\,{\sc iii}
and Fe\,{\sc iii} lines. The effective temperature estimated from
six C\,{\sc ii} lines spanning excitation potentials from 16 eV
to 23 eV is $T_{\rm eff} = 16100\pm300$ K.
The $\log g$ was found by
combining this estimate of $T_{\rm eff}$ with loci from ionization
equilibrium in the $\log g$ vs $T_{\rm eff}$ plane (Figure 7). Loci were
provided by C\,{\sc ii}/C\,{\sc iii}, Si\,{\sc ii}/Si\,{\sc iii},
Mn\,{\sc ii}/Mn\,{\sc iii}, Fe\,{\sc ii}/Fe\,{\sc iii}, 
Co\,{\sc ii}/Co\,{\sc iii}, and Ni\,{\sc ii}/Ni\,{\sc iii}. 
The weighted mean estimate is $\log g = 2.3\pm0.25$ cgs.
Results of the abundance analysis with a STERNE model corresponding
to (16100, 2.3, 10) are given in Table 7. 
The $v\sin i$ is deduced to be about 4 km s$^{-1}$.

\subsubsection{The optical spectrum}

Sch\"{o}nberner \& Wolf's (1974) analysis was undertaken with
an unblanketed model atmosphere corresponding to (16000, 2.2, 10).
\citet{heb83} revised the 1974 abundances using a blanketed model
corresponding to (15500, 2.1, 10). Here, Sch\"{o}nberner \&
Wolf's list of lines and their equivalent width have been
reanalysed using our $gf$-values and a microturbulent velocity of 12 km s$^{-1}$
found from the N\,{\sc ii} lines. Two sets of  model atmosphere
parameters  are considered: Heber's (1983) and ours from the  {\it STIS} spectrum.
Results are given in Table 7.

This EHe was observed with the Vainu Bappu Telescope's
fiber-fed cross-dispersed echelle spectrograph.
Key lines were identified across the observed limited 
wavelength regions. 
The microturbulent velocity is judged to be about 12 km s$^{-1}$ from weak and 
strong lines of N\,{\sc ii} and S\,{\sc ii}. 
The effective temperature estimated from
seven C\,{\sc ii} lines spanning excitation potentials from 14 eV
to 23 eV is $T_{\rm eff} = 15500\pm500$ K.
The wings of the observed He\,{\sc i}
profile at 6678.15\AA\ are used to determine the surface gravity.
The He\,{\sc i} profile is best reproduced by  $\log g = 1.9\pm0.2$ cgs 
for the derived $T_{\rm eff}$ of 15500 K.
Hence, the model atmosphere (15500, 1.9, 12) is
adopted to derive the abundances given in Table 7. 

\clearpage
\begin{figure}
\epsscale{1.00}
\plotone{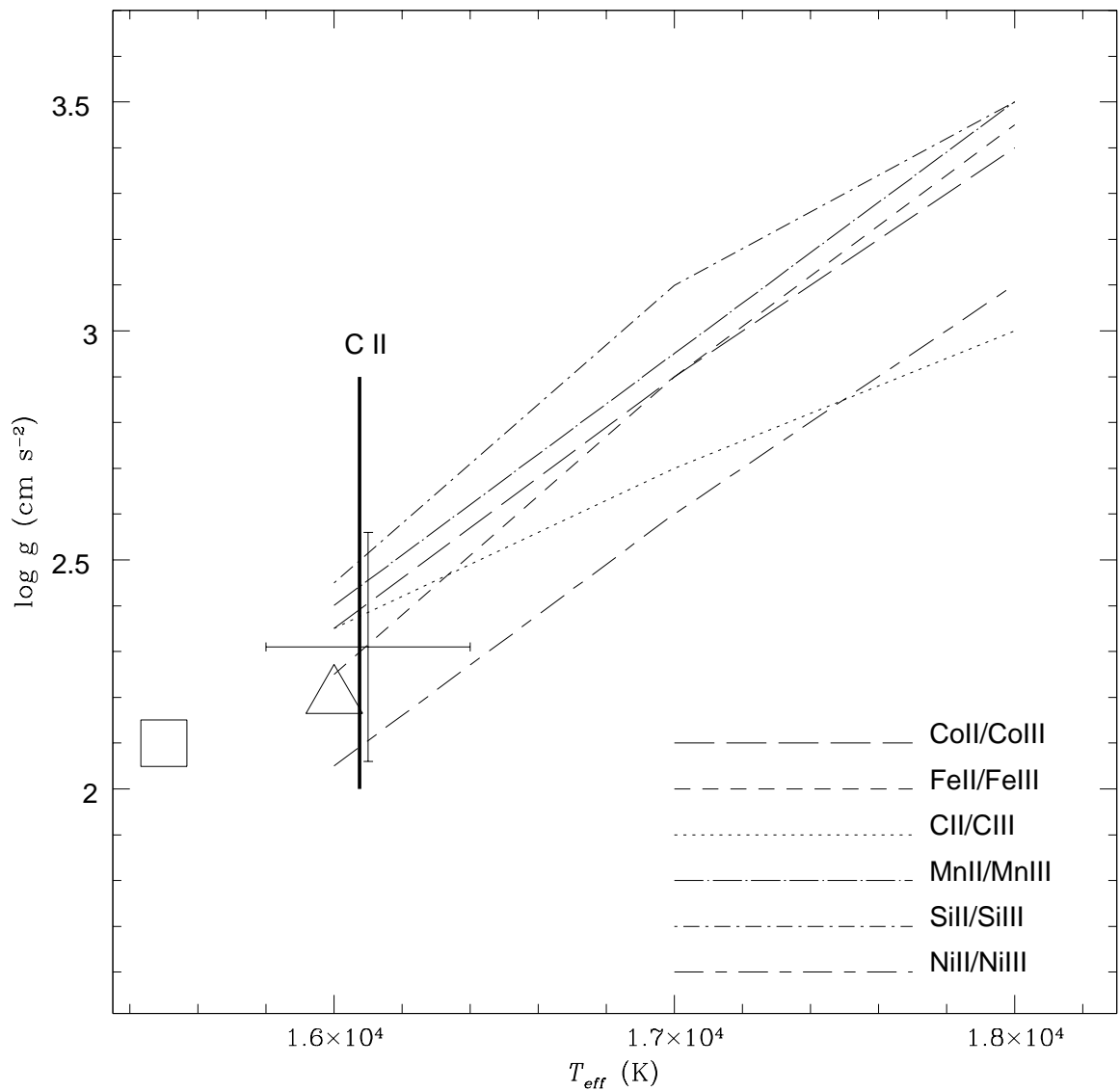}
\caption{The $T_{\rm eff}$ vs $\log g$ plane for HD\,124448 from
analysis of the {\it STIS} spectrum.
Ionization equilibria are plotted  -- see keys on the figure.
The locus satisfying the excitation balance of C\,{\sc ii}
lines is shown by thick solid line.
The cross shows the adopted model atmosphere parameters.
The large triangle shows the parameters chosen by \citet{sch74}
from their analysis of an optical spectrum using
unblanketed model atmospheres.
The large square shows the revised parameters by \citet{heb83} using
blanketed model atmospheres.
\label{fig7}} 
\end{figure}
\clearpage

\subsubsection{Adopted abundances}

Adopted abundances are given in Table 4. These are based on our
$STIS$ and optical ($VBT$) spectra. If the key lines are not available
in the $STIS$ and optical ($VBT$) spectra, then the abundances are
from Sch\"{o}nberner \& Wolf's list of lines and their equivalent width
using our $gf$-values and the model (16100, 2.3, 12).
Our limit on the H abundance is from
the absence of the H$\alpha$ line in the VBT spectrum.
The C abundance is from ultraviolet C\,{\sc ii} and C\,{\sc iii} lines,
and optical C\,{\sc ii} lines weighted by the number of lines.
The N abundance is from the optical N\,{\sc ii} lines because the
ultraviolet N\,{\sc ii} lines are blended. The O abundance is from the optical
O\,{\sc ii} lines from Sch\"{o}nberner \& Wolf's list.
The Mg abundance is from the ultraviolet and the optical Mg\,{\sc ii} lines
which are given equal weight and are weighted by their numbers.
The ultraviolet Al\,{\sc iii} (saturated and blended)
line is given least weight. 
The Al abundance is from the ultraviolet and optical Al\,{\sc ii} lines, and 
optical Al\,{\sc iii} lines weighted by the number of lines.
Equal weight is given to ultraviolet and optical Si\,{\sc ii} lines, and
the adopted Si abundance from Si\,{\sc ii} lines is weighted by the 
number of lines. The mean Si abundance from ultraviolet and optical 
Si\,{\sc iii} lines is consistent with the adopted Si abundance from 
Si\,{\sc ii} lines. The S abundance is from the optical S\,{\sc ii} lines
($VBT$ spectrum) and is found consistent with the S\,{\sc ii} and S\,{\sc iii}
lines from Sch\"{o}nberner \& Wolf's list for our  $gf$-values.
The Fe abundance is from ultraviolet Fe\,{\sc ii} and Fe\,{\sc iii}
lines weighted by the number of lines.
Ni abundance is from ultraviolet Ni\,{\sc ii} and Ni\,{\sc iii} lines
weighted by the number of lines.

\clearpage

\begin{table}
\begin{center}\small{Table 7} \\
\small{Chemical Composition of the EHe HD\,124448}\\
\begin{tabular}{lrccccrccccl}
\hline
\hline
& & UV$^{\rm a}$ & & &  &  & & & Optical  & & \\
\cline{2-4}\cline{7-12}
Species &  log $\epsilon$ & & $n$\ \ \ & & &   log $\epsilon ^{\rm b}$ 
& log $\epsilon ^{\rm c}$ & log $\epsilon ^{\rm d}$ & $n$ & log $\epsilon ^{\rm e}$ & $n$ \\
\hline
H\,{\sc i} & \nodata & & \nodata & & &  $< 7.5$ & $< 7.5$ & $< 7.5$ &2 & $< 6.3$ & 1\\
He\,{\sc i} & \nodata & & \nodata \ \ & & &  11.53 & 11.53 & 11.53 &\nodata & 11.53 &\nodata\\
C\,{\sc ii} & 9.3 && 8  \ \ &&& 9.0 & 9.0 & 9.5 & 7 & 9.1 & 7\\
C\,{\sc iii} & 9.2 && 2 \ \  &&& \nodata & \nodata & \nodata & \nodata & \nodata & \nodata\\
N\,{\sc ii} & 8.8:&& 3 \ \  &&& 8.4 & 8.4 & 8.8 &18 & 8.6 & 3\\
O\,{\sc ii} & \nodata && \nodata &&& 8.1 & 8.1 & 8.5 & 5 & \nodata & \nodata\\
Mg\,{\sc ii} & 7.5 &&  2 \ \  &&& 8.3 & 8.3 & 8.2 & 2 & 7.9 & 1\\
Al\,{\sc ii} & 6.3 && 1 \ \  &&& 6.3 & 6.3 & 6.3 & 1 & 6.6 & 2\\
Al\,{\sc iii} & 6.1: && 1 \ \ &&& 5.6 & 5.6 & 5.9 & 1 & 6.5 & 1\\
Si\,{\sc ii} & 7.2 && 3 \ \ &&& 7.1 & 7.2 & 7.6 & 3 & 6.9 & 1\\
Si\,{\sc iii} & 6.9 && 1 \ \ &&& 6.7 & 6.7 & 7.3 & 6 & 7.5 & 1\\
P\,{\sc iii} & \nodata && \nodata &&& 5.2 & 5.2 & 5.6 & 2 & \nodata & \nodata\\
S\,{\sc ii} & \nodata && \nodata &&& 7.0 & 7.0 & 7.0 & 9 & 6.9 & 3\\
S\,{\sc iii} & \nodata && \nodata &&& 6.9 & 6.9 & 7.3 & 4 & \nodata & \nodata\\
Ar\,{\sc ii} & \nodata && \nodata &&& 6.5 & 6.5 & 6.6 & 3 & \nodata & \nodata\\
Ca\,{\sc ii} & \nodata && \nodata &&& $<$6.1 & $<$6.0 & $<$6.9 & 2 & \nodata & \nodata\\
Ti\,{\sc ii} & \nodata && \nodata \ \ &&& 6.1 & 6.2 & 5.1 & 3 & \nodata & \nodata\\
Ti\,{\sc iii} & 4.8 && 1 \ \ &&& \nodata & \nodata & \nodata & \nodata & \nodata & \nodata\\
Cr\,{\sc iii} & 5.2 && 19 \ \ &&& \nodata & \nodata & \nodata & \nodata & \nodata & \nodata\\
Mn\,{\sc ii} & 4.9 && 3 \ \ &&& \nodata & \nodata & \nodata & \nodata & \nodata & \nodata\\
Mn\,{\sc iii} & 4.9 && 6 \ \ &&& \nodata & \nodata & \nodata & \nodata & \nodata & \nodata\\
Fe\,{\sc ii} & 7.2 && 21 \ \ &&& 7.5 & 7.7 & 7.8 & 4& \nodata & \nodata\\
Fe\,{\sc iii} & 7.2 && 9 \ \ &&& \nodata & \nodata & \nodata & \nodata & \nodata & \nodata\\
Co\,{\sc ii} & 4.6 && 4 \ \ &&& \nodata & \nodata & \nodata & \nodata & \nodata & \nodata\\
Co\,{\sc iii} & 4.6 && 3 \ \ &&& \nodata & \nodata & \nodata & \nodata & \nodata & \nodata\\
Ni\,{\sc ii}    & 5.6 & & 26 \ \ &&& \nodata & \nodata & \nodata & \nodata & \nodata & \nodata\\
Ni\,{\sc iii} & 5.8 && 3 \ \ &&& \nodata & \nodata & \nodata & \nodata & \nodata & \nodata\\
Y\,{\sc iii} & 2.2 && 2 \ \ &&& \nodata & \nodata & \nodata & \nodata & \nodata & \nodata\\
Zr\,{\sc iii} & 2.7 && 3 \ \ &&& \nodata & \nodata & \nodata & \nodata & \nodata & \nodata\\
Ce\,{\sc iii} & $< 1.8$ && 1 \ \ &&& \nodata & \nodata & \nodata & \nodata & \nodata & \nodata\\
\hline
\end{tabular}
\end{center}
$^{\rm a}$ This paper from the model atmosphere (16100, 2.3, 10.0)\\
$^{\rm b}$ Rederived from the list of \citet{sch74} and Heber's (1983) revised model
parameters (15500, 2.1, 12.0)\\
$^{\rm c}$ Our results from Sch\"{o}nberner \& Wolf's line lists and our
 $STIS$-based model atmosphere (16100, 2.3, 12.0)\\
$^{\rm d}$ From \citet{sch74}\\
$^{\rm e}$ Abundances from the  $VBT$ echelle spectrum and the model
atmosphere (15500, 1.9, 12.0)\\
\end{table}

\clearpage

\subsection{PV\,Tel = HD\,168476}

This star was discovered by \citet{tha52}
in a southern hemisphere survey of high galactic latitude B stars
following Popper's discovery of HD\,124448. The
star's chemical composition was determined via a model atmosphere
by \citet{walk81}, see also \citet{heb83},
from photographic optical spectra.

\subsubsection{The ultraviolet spectrum}

A microturbulent velocity of 9$\pm1$ km s$^{-1}$ is found from Cr\,{\sc iii}
and Fe\,{\sc iii} lines.
 The effective temperature estimated from
Fe\,{\sc ii} lines spanning excitation potentials from 0 eV
to 9 ev is $T_{\rm eff} = 13500\pm500$ K.
 The effective temperature estimated from
Ni\,{\sc ii} lines spanning about 8 eV in excitation potential
is $T_{\rm eff} = 14000\pm500$ K. We adopt $T_{\rm eff} = 13750\pm400$ K.
 The $\log g$ was found by
combining this estimate of $T_{\rm eff}$ with loci from ionization
equilibrium in the $\log g$ vs $T_{\rm eff}$ plane (Figure 8). Loci were
provided by C\,{\sc ii}/C\,{\sc iii}, Cr\,{\sc ii}/Cr\,{\sc iii},
Mn\,{\sc ii}/Mn\,{\sc iii}, and Fe\,{\sc ii}/Fe\,{\sc iii}.
The mean estimate is $\log g = 1.6\pm0.25$ cgs.
The $v\sin i$ is deduced to be about 25 km s$^{-1}$.
Results of the abundance analysis with a STERNE model corresponding 
to (13750, 1.6, 9) are given in Table 8.

\subsubsection{The optical spectrum}

\citet{walk81} analysis was undertaken with
an unblanketed model atmosphere corresponding to (14000, 1.5, 10).
\citet{heb83} reconsidered the 1981 abundances using a blanketed model
corresponding to (13700, 1.35, 10), a model with parameters
very similar to our UV-based results.
Here, Walker \& Sch\"{o}nberner's
list of lines and their equivalent width have been
reanalysed using our $gf$-values.
The microturbulent velocity of 15$\pm4$ km s$^{-1}$ was found from
the N\,{\sc ii} lines, and S\,{\sc ii} lines.
The $T_{\rm eff}$ and $\log g$ were taken from the {\it STIS}
analysis.  
Results are given in Table 8. Several elements considered by
Walker \& Sch\"{o}nberner are omitted here because their
lines give a large scatter, particularly for lines with wavelengths
shorter than about 4500\AA.
\clearpage
\begin{figure}
\epsscale{1.00}
\plotone{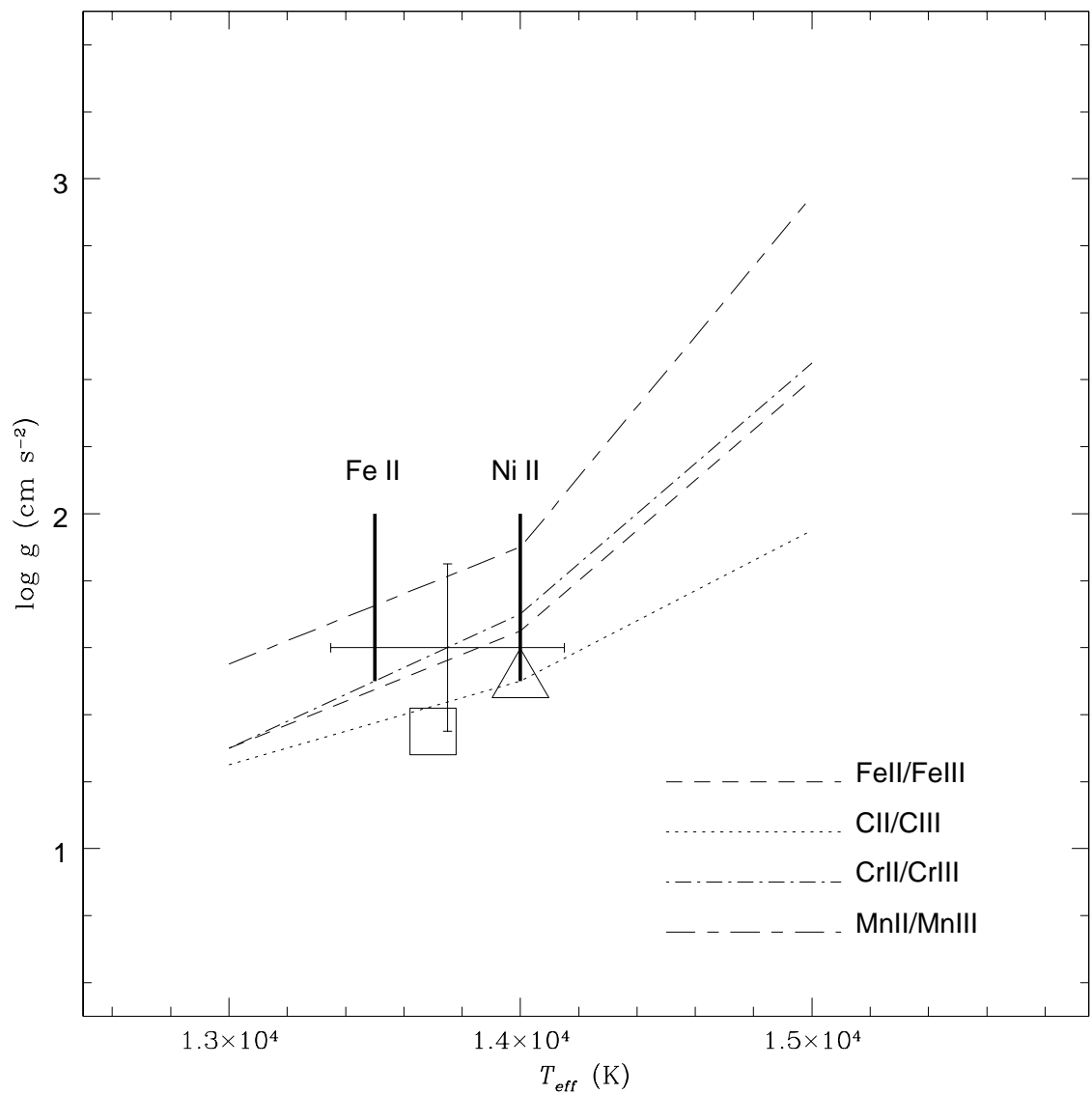}
\caption{The $T_{\rm eff}$ vs $\log g$ plane for PV\,Tel from
analysis of the {\it STIS} spectrum.
Ionization equilibria are plotted  -- see keys on the figure.
The loci satisfying the excitation balance of Fe\,{\sc ii}
and Ni\,{\sc ii} lines are shown by thick solid lines.
The cross shows the adopted model atmosphere parameters.
The large triangle shows the parameters chosen by 
\citet{walk81} from their analysis of an optical 
spectrum using unblanketed model atmospheres.
The large square shows the revised parameters by \citet{heb83} using
blanketed model atmospheres.
\label{fig8}} 
\end{figure}
\clearpage

\subsubsection{Adopted abundances}

Adopted abundances are given in Table 4.
The C abundances from ultraviolet and optical C\,{\sc ii} lines agree well.
More weight is given to C\,{\sc ii} lines over ultraviolet C\,{\sc iii} line.
The N abundance from optical N\,{\sc ii} lines is about 8.6$\pm$0.2,
a reasonable standard deviation. Adopted N abundances
are from optical N\,{\sc ii} lines. 
The O abundance is from O\,{\sc i} lines in the optical red region.
The Mg abundance is from optical Mg\,{\sc ii} lines.
The Al abundance is uncertain; the several Al\,{\sc ii} and
Al\,{\sc iii} lines do not yield very consistent results.
The Al\,{\sc iii} line at 4529.20 \AA\ gives an Al abundance 
(6.2)  which is close to that derived (6.1) from the ultraviolet Al\,{\sc iii} line.
More weight is given to optical Si lines than the ultraviolet Si\,{\sc ii}
lines. The adopted Si abundance is the simple mean of
optical Si\,{\sc ii} and Si\,{\sc iii} lines.
The P abundance is from the optical red P\,{\sc ii} lines. 
The S abundance is from S\,{\sc ii} and S\,{\sc iii} optical lines.
Adopted Cr abundance is from ultraviolet Cr\,{\sc ii} and 
Cr\,{\sc iii} lines weighted by the number of lines.
Adopted Mn abundance is from
ultraviolet Mn\,{\sc ii} and Mn\,{\sc iii} lines weighted by their number.
Adopted Fe abundance is from ultraviolet Fe\,{\sc ii} and Fe\,{\sc iii} lines.
Adopted Ni abundance is from ultraviolet Ni\,{\sc ii}  lines.

\clearpage
\begin{table}
\begin{center}\small{Table 8} \\
\small{Chemical Composition of the EHe PV\,Tel}\\
\begin{tabular}{lrccccrcl}
\hline
\hline
& & UV$^{\rm a}$ & & &  &  & Optical $^{\rm b}$ & \\
\cline{2-4}\cline{7-9}
Species &  log $\epsilon$ & & $n$\ \ \ & & &   log $\epsilon$ &  & $n$  \\
\hline
H\,{\sc i} & \nodata & & \nodata & & &  $< 7.3$ & &2 \\
He\,{\sc i} & \nodata & & \nodata & & &  \nodata & &\nodata \\
C\,{\sc ii} & 9.3 && 2 \ \ &&& 9.3 && 2 \\
C\,{\sc iii} & 9.6 && 1 \ \  &&& \nodata && \nodata \\
N\,{\sc ii} & \nodata&& \nodata \ \  &&& 8.6 & &19 \\
O\,{\sc i} & \nodata && \nodata &&& 8.6 && 2\\
O\,{\sc ii} & \nodata && \nodata &&& 8.1 && 1\\
Mg\,{\sc ii} & 8.0: &&  1 \ \  &&& 7.8 && 10\\
Al\,{\sc ii} & \nodata && \nodata  &&& 7.5 && 2 \\
Al\,{\sc iii} & 6.1 && 1 \ \ &&& 6.6 && 2\\
Si\,{\sc ii} & 6.8 && 2 \ \ &&& 7.5 && 6\\
Si\,{\sc iii} & \nodata && \nodata &&& 7.1 && 3\\
P\,{\sc ii} & \nodata && \nodata &&& 6.1 && 4\\
S\,{\sc ii} & \nodata && \nodata &&& 7.2 && 45\\
S\,{\sc iii} & \nodata && \nodata &&& 7.2 && 4\\
Ti\,{\sc ii} & 5.2 && 2 \ \ &&& \nodata && \nodata\\
Cr\,{\sc ii} & 5.0 && 2 \ \ &&& \nodata && \nodata \\
Cr\,{\sc iii} & 5.1 && 16 \ \ &&& \nodata && \nodata\\
Mn\,{\sc ii} & 5.1 && 2 \ \ &&& \nodata && \nodata \\
Mn\,{\sc iii} & 4.8 && 4 \ \ &&& \nodata && \nodata \\
Fe\,{\sc ii} & 7.0 && 24 \ \ &&& \nodata && \nodata\\
Fe\,{\sc iii} & 7.1 && 11 \ \ &&& \nodata && \nodata\\
Ni\,{\sc ii}    & 5.7 & & 16 \ \ &&& \nodata && \nodata\\
Y\,{\sc iii} & 2.9 && 1 \ \ &&& \nodata && \nodata\\
Zr\,{\sc iii} & 3.1 && 4 \ \ &&& \nodata && \nodata\\
Ce\,{\sc iii} & $< 1.7$ && 1 \ \ &&& \nodata && \nodata\\
\hline
\end{tabular}
\end{center}
$^{\rm a}$  This paper and the model atmosphere (13750, 1.6, 9.0)\\
$^{\rm b}$ Recalculation of Walker \& Sch\"{o}nberner's (1981) line list
using the model atmosphere (13750, 1.6, 15.0)\\
\end{table}
\clearpage

\subsection{V2244 Ophiuchi = LS\,IV-1$^{\circ}$\,2}

\subsubsection{The ultraviolet spectrum}

The UV spectrum of V2244 Oph is of poor quality owing to an
inadequate exposure time. This line-rich spectrum is usable
only at wavelengths longer than about 2200 \AA.
Given the low S/N ratio over a restricted wavelength
interval, we did not attempt to derive the atmospheric
parameters from the UV spectrum but adopted the values
obtained earlier from a full analysis of a high-quality
optical spectrum \citep{pan01}: $T_{\rm eff} = 12,750$ K,
$\log g = 1.75$ cgs, and $\xi = 10$ km s$^{-1}$.
Abundances derived from the UV spectrum are given in Table 9
with results from \citet{pan01} from a high-quality
optical spectrum. 

\subsubsection{Adopted abundances}

For the few ions with UV and optical lines, the abundances are 
in good agreement. Adopted abundances are given in Table 4. 
More weight is given to the optical lines over UV lines because
UV lines are not very clean.

\clearpage
\begin{table}
\begin{center}\small{Table 9} \\
\small{Chemical Composition of the EHe LS\,IV-1$^{\circ}$\,2}\\
\begin{tabular}{lrccccrcl}
\hline
\hline
& & UV$^{\rm a}$ & & &  &  & Optical $^{\rm b}$ & \\
\cline{2-4}\cline{7-9}
Species &  log $\epsilon$ & & $n$\ \ \ & & &   log $\epsilon$ &  & $n$  \\
\hline
H\,{\sc i} & \nodata & & \nodata & & &  7.1 & &1 \\
He\,{\sc i} & \nodata & & \nodata & & &  11.54 & & 1\\
C\,{\sc i} & \nodata && \nodata \ \ &&& 9.3 && 15 \\
C\,{\sc ii} & 9.5 && 2 \ \ &&& 9.3 && 7 \\
N\,{\sc i} & \nodata && \nodata \ \  &&& 8.2 & &6 \\
N\,{\sc ii} & \nodata && \nodata \ \  &&& 8.3 & &14 \\
O\,{\sc i} & \nodata && \nodata &&& 8.8 && 3\\
O\,{\sc ii} & \nodata && \nodata &&& 8.9 && 5\\
Mg\,{\sc ii} & 6.9 &&  1 \ \  &&& 6.9 && 6\\
Al\,{\sc ii} & \nodata && \nodata  &&& 5.4 && 8 \\
Si\,{\sc ii} & 6.2 && 1 \ \ &&& 5.9 && 3\\
P\,{\sc ii} & \nodata && \nodata &&& 5.1 && 3\\
S\,{\sc ii} & \nodata && \nodata &&& 6.7 && 35\\
Ca\,{\sc ii} & \nodata && \nodata &&& 5.8 && 2\\
Ti\,{\sc ii} & \nodata && \nodata \ \ &&& 4.7 && 5\\
Cr\,{\sc iii} & 5.0 && 3 \ \ &&& \nodata && \nodata\\
Fe\,{\sc ii} & 6.2 && 6 \ \ &&& 6.3 && 22\\
Fe\,{\sc iii} & \nodata && \nodata \ \ &&& 6.1 && 2\\
Ni\,{\sc ii}    & 5.1 & & 3 \ \ &&& \nodata && \nodata\\
Y\,{\sc iii} & 1.4 && 1 \ \ &&& \nodata && \nodata\\
Zr\,{\sc iii} & 2.3 && 3 \ \ &&& \nodata && \nodata\\
Nd\,{\sc iii} & \nodata && \nodata \ \ &&& $<$0.8 && 2\\
\hline
\end{tabular}
\end{center}
$^{\rm a}$ Derived using Pandey et al.'s (2001) model
atmosphere (12750, 1.75, 10.0)\\
$^{\rm b}$ Taken from \citet{pan01}. Their analysis
uses the model atmosphere (12750, 1.75, 10.0)\\
\end{table}
\clearpage
\subsection{FQ\,Aquarii}

\subsubsection{The ultraviolet spectrum}

A microturbulent velocity of 7.5$\pm1.0$ km s$^{-1}$ is provided by the
Cr\,{\sc ii} and Fe\,{\sc ii} lines. The Fe\,{\sc ii} lines
spanning about 7 eV in excitation potential suggest that
$T_{\rm eff} = 8750\pm300$ K. This temperature in
conjunction with the ionization equilibrium loci for
Si\,{\sc i}/Si\,{\sc ii}, Cr\,{\sc ii}/Cr\,{\sc iii},
Mn\,{\sc ii}/Mn\,{\sc iii}, and Fe\,{\sc ii}/Fe\,{\sc iii}
gives the surface gravity $\log g = 0.3\pm0.3$ cgs (Figure 9). 
The $v\sin i$ is deduced to be about 20 km s$^{-1}$.
Abundances are given in Table 10 for the model
corresponding to (8750, 0.3, 7.5)
along with the abundances obtained from an optical spectrum
by \citet{pan01}. A model corresponding to (8750, 0.75, 7.5)
was used by Pandey et al. which is very similar to
our UV-based results.
\clearpage
\begin{figure}
\epsscale{1.00}
\plotone{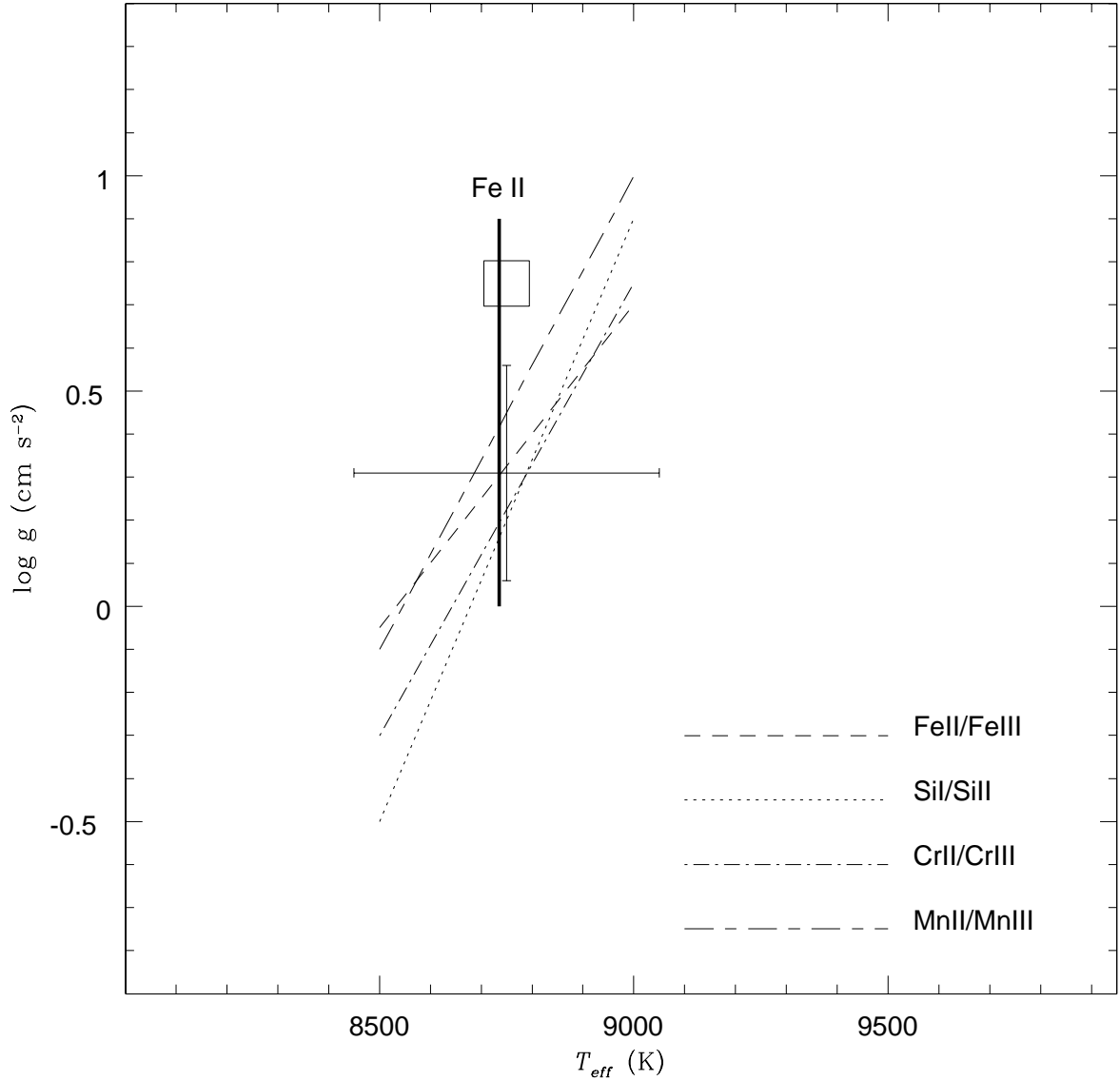}
\caption{The $T_{\rm eff}$ vs $\log g$ plane for FQ\,Aqr from
analysis of the {\it STIS} spectrum.
Ionization equilibria are plotted  -- see keys on the figure.
The locus satisfying the excitation balance of Fe\,{\sc ii}
lines is shown by thick solid line.
The cross shows the adopted model
atmosphere parameters. The large square shows the parameters
chosen by \citet{pan01} from their analysis of an optical spectrum.
\label{fig9}}
\end{figure}
\clearpage

\subsubsection{Adopted abundances}

Adopted abundances are given in Table 4. 
The C and N abundances are from \citet{pan01}
because the ultraviolet  C\,{\sc ii},  C\,{\sc iii}  and N\,{\sc ii}
are blended.
The ultraviolet Al\,{\sc ii} line is blended and 
is given no weight.
Equal weight is given to ultraviolet Si\,{\sc i} and Si\,{\sc ii} lines,
and Pandey et al.'s Si abundance. The mean Si abundance is from these lines
weighted by the number of lines.
Ca abundance is from Pandey et al. 
Equal weight is given to ultraviolet Cr\,{\sc ii} and Cr\,{\sc iii} lines,
which give a Cr abundance in good agreement with Pandey et al.
Equal weight is given to the abundances based on ultraviolet Mn\,{\sc ii} lines 
and Pandey et al.'s Mn abundance. No weight is given to ultraviolet Mn\,{\sc iii}
lines because they are blended. 
 A simple mean of ultraviolet and optical based Mn abundance
is adopted.
The Fe abundance is from ultraviolet Fe\,{\sc ii}, Fe\,{\sc iii}, and
Pandey et al.'s optical Fe\,{\sc i}, Fe\,{\sc ii} lines weighted by the number of lines.
The Zr abundance from ultraviolet Zr\,{\sc iii} lines is in  agreement with 
Pandey et al.'s optical Zr\,{\sc ii} lines within the expected uncertainties.
The adopted Zr abundance is a simple mean of ultraviolet and optical 
based Zr abundances.

\clearpage
\begin{table}
\begin{center}
\small{Table 10} \\
\small{Chemical Composition of the EHe FQ\,Aqr}\\
\begin{tabular}{lrccccrcl}
\hline
\hline
& & UV$^{\rm a}$ & & &  &  & Optical $^{\rm b}$ & \\
\cline{2-4}\cline{7-9}
Species &  log $\epsilon$ & & $n$\ \ \ & & &   log $\epsilon$ &  & $n$  \\
\hline
H\,{\sc i} & \nodata & & \nodata & & &  6.2 & &1 \\
He\,{\sc i} & \nodata & & \nodata & & &  11.54 & & 3\\
C\,{\sc i} & \nodata && \nodata \ \ &&& 9.0 && 30 \\
C\,{\sc ii} & 9.3: && 1 \ \ &&& 9.0 && 2 \\
N\,{\sc i} & \nodata && \nodata \ \  &&& 7.1 & &5 \\
N\,{\sc ii} & 6.7: && 1 \ \  &&& 7.2 & &2 \\
O\,{\sc i} & \nodata && \nodata &&& 8.9 && 8\\
Mg\,{\sc i} & \nodata &&  \nodata \ \  &&& 5.5 && 5\\
Mg\,{\sc ii} & 6.0 &&  1 \ \  &&& 6.0 && 6\\
Al\,{\sc ii} & 4.7: && 1  &&& 4.7 && 4 \\
Si\,{\sc i} & 6.0 && 6 \ \ &&& \nodata && \nodata\\
Si\,{\sc ii} & 6.0 && 3 \ \ &&& 6.3 && 6\\
P\,{\sc ii} & \nodata && \nodata &&& 4.2 && 2\\
S\,{\sc i} & \nodata && \nodata &&& 6.1 && 3\\
S\,{\sc ii} & \nodata && \nodata &&& 5.9 && 7\\
Ca\,{\sc i} & \nodata && \nodata &&& 4.0 && 1\\
Ca\,{\sc ii} & 4.3: && 2 &&& 4.2 && 1\\
Sc\,{\sc ii} & \nodata && \nodata &&& 2.1 && 7\\
Ti\,{\sc ii} & \nodata && \nodata \ \ &&& 3.2 && 42\\
Cr\,{\sc ii} & 3.6 && 11 \ \ &&& 3.6 && 30 \\
Cr\,{\sc iii} & 3.6 && 5 \ \ &&& \nodata && \nodata\\
Mn\,{\sc ii} & 3.5 && 3 \ \ &&& 4.3 && 3 \\
Mn\,{\sc iii} & 3.5: && 3 \ \ &&& \nodata && \nodata \\
Fe\,{\sc i} & \nodata && \nodata \ \ &&& 5.1 && 7\\
Fe\,{\sc ii} & 5.5 && 25 \ \ &&& 5.4 && 59\\
Fe\,{\sc iii} & 5.4 && 11 \ \ &&& \nodata && \nodata\\
Co\,{\sc ii} & 3.0 && 4 \ \ &&& \nodata && \nodata\\
Ni\,{\sc ii}    & 4.0 & & 7 \ \ &&& \nodata && \nodata\\
Cu\,{\sc ii}    & 2.7 & & 4 \ \ &&& \nodata && \nodata\\
Zn\,{\sc ii}    & 3.2 & & 2 \ \ &&& \nodata && \nodata\\
Zr\,{\sc ii} & \nodata && \nodata \ \ &&& 0.8 && 2\\
Zr\,{\sc iii} & 1.1 && 6 \ \ &&& \nodata && \nodata\\
Ce\,{\sc iii} & $< 0.3$ && 1 \ \ &&& \nodata && \nodata\\
\hline
\end{tabular}
\end{center}
$^{\rm a}$  This paper from the {\it STIS} spectrum using the model
atmosphere (8750, 0.3, 7.5) \\
$^{\rm b}$ See \citet{pan01}. Their analysis uses the
model atmosphere (8750, 0.75, 7.5) \\
\end{table}
\clearpage
%
%

\section{Abundances - clues to EHes origin and evolution}

In this section, we examine correlations between the
abundances measured for the EHes.
It has long been considered that the atmospheric
composition of an EHe is at the least a blend of
the star's original composition, material exposed
to H-burning reactions, and to products from layers
in which He-burning has occurred. We comment on
the abundance correlations with this minimum
model in mind.
This section is followed
by discussion on the abundances in light of the scenario of a merger of
two white dwarfs.

Our sample of seven EHes (Table 4) is augmented by results from
the literature for an additional ten EHes. These
range in effective temperature from the hottest at 32,000 K
to the coolest at 9500 K. (The  temperature range of our septet is 18300 K
to 8750 K.) From hottest to coolest, the additional stars are
LS IV $+6^\circ2$ \citep{jeff98}, V652\,Her \citep{jeff99},
LSS\,3184 \citep{drill98}, HD\,144941 \citep{har97,jeff97},
BD$-9^\circ4395$ \citep{jeff92}, DY\,Cen (Jeffery \&
Heber 1993), LSS\,4357 and LSS\,99 \citep{jef98}, and
LS IV $-14^\circ109$ and BD$-1^\circ3438$ \citep{pan01}.
DY\,Cen might be more properly regarded as
a hot R Corona Borealis (RCB) variable. As a reference 
mixture, we have adopted the solar abundances from 
Table 2 of \citet{lod03} (see Table 4).

\subsection{Initial metallicity}

The initial metallicity for an EHe composition is the abundance (i.e., mass fraction)
of an element unlikely to be affected by H- and He-burning and
attendant nuclear reactions. We take Fe as our initial
choice for the representative of initial metallicity, and examine first the
correlations between Cr, Mn, and Ni, three elements with
reliable abundances uniquely or almost so provided
from the {\it STIS} spectra. Data are included for two cool EHes analysed by 
\citet{pan01} from optical spectra alone. Figure 10
shows that Cr, Mn, and Ni vary in concert, as expected.
An apparently discrepant star with a high Ni abundance
is the cool EHe LS IV $-14^\circ 109$
from \citet{pan01}, but the Cr and Mn abundances
are as expected.
\clearpage
\begin{figure}
\epsscale{1.00}
\plotone{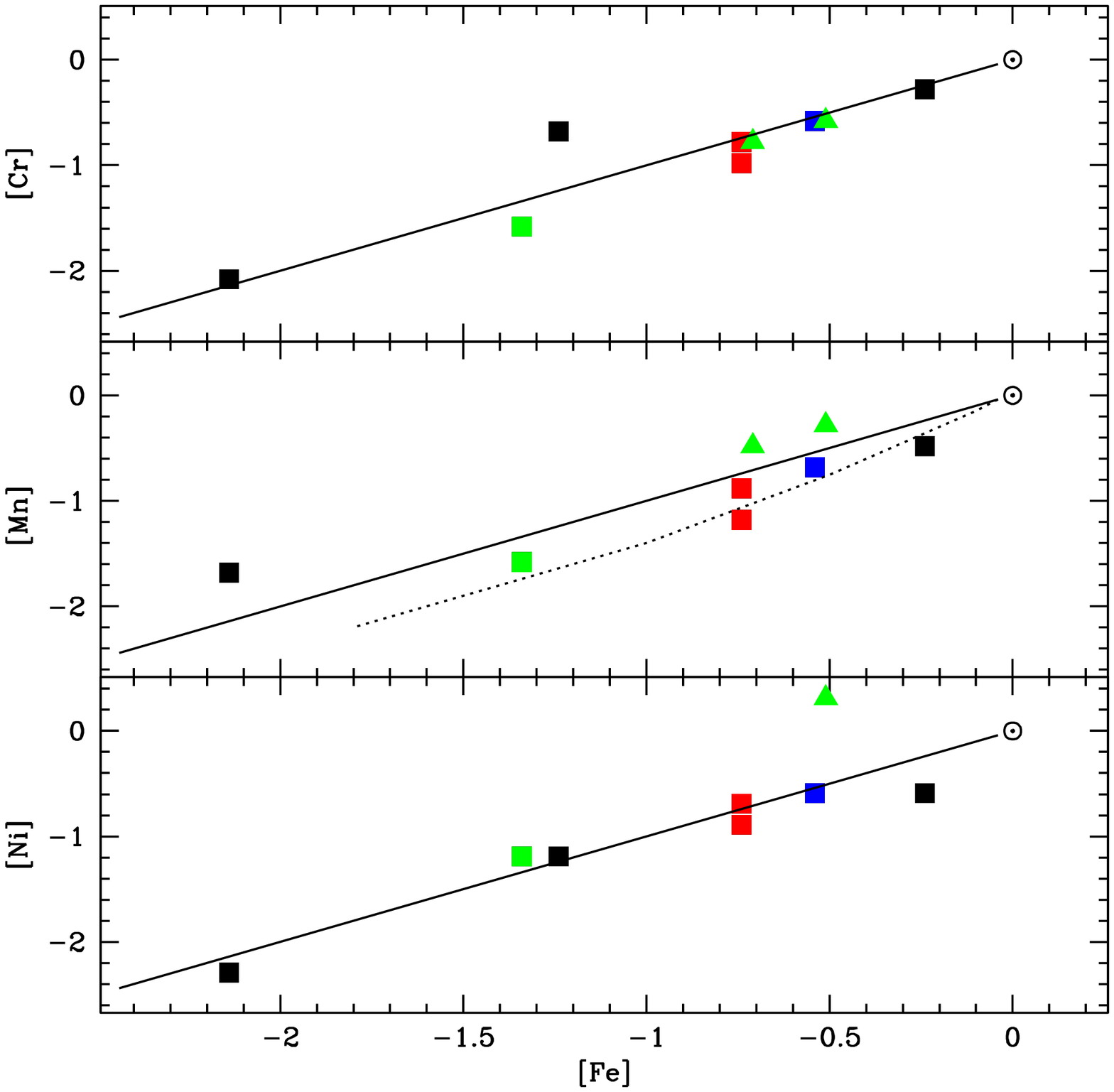}
\caption{[Cr], [Mn], and [Ni] vs [Fe]. Our sample of seven EHes
is represented by filled squares. Two cool EHes analysed by
\citet{pan01} are represented by filled triangles.
$\odot$ represents Sun. [X] = [Fe] are denoted by the solid 
lines where X represents Cr, Mn, and Ni. The dotted line for Mn is
from the relation [Mn/Fe] versus [Fe/H] for normal
disk and halo stars given by B.E. Reddy (private communication) and
\citet{red03}.
\label{fig10}}
\end{figure}
\clearpage
A second group of elements expected to be unaffected or only
slightly so by nuclear reactions associated with H- and He-burning
is the $\alpha$-elements Mg, Si, S, and Ca and also Ti.
The variation of these abundances with the Fe abundance
is shown in Figure  11 together with a mean (denoted
by $\alpha$) computed from the abundances of Mg, Si, and S.
It is known that in metal-poor normal and unevolved stars that the
abundance ratio $\alpha$/Fe varies with Fe \citep{ryd04,gos00}. 
This variation is characterized by the dotted line in the figure. 
Examination of Figure 11 suggests that the abundances of the
$\alpha$-elements and Ti follow the expected trend with the
dramatic exception of DY\,Cen.
\clearpage
\begin{figure}
\epsscale{1.00}
\plotone{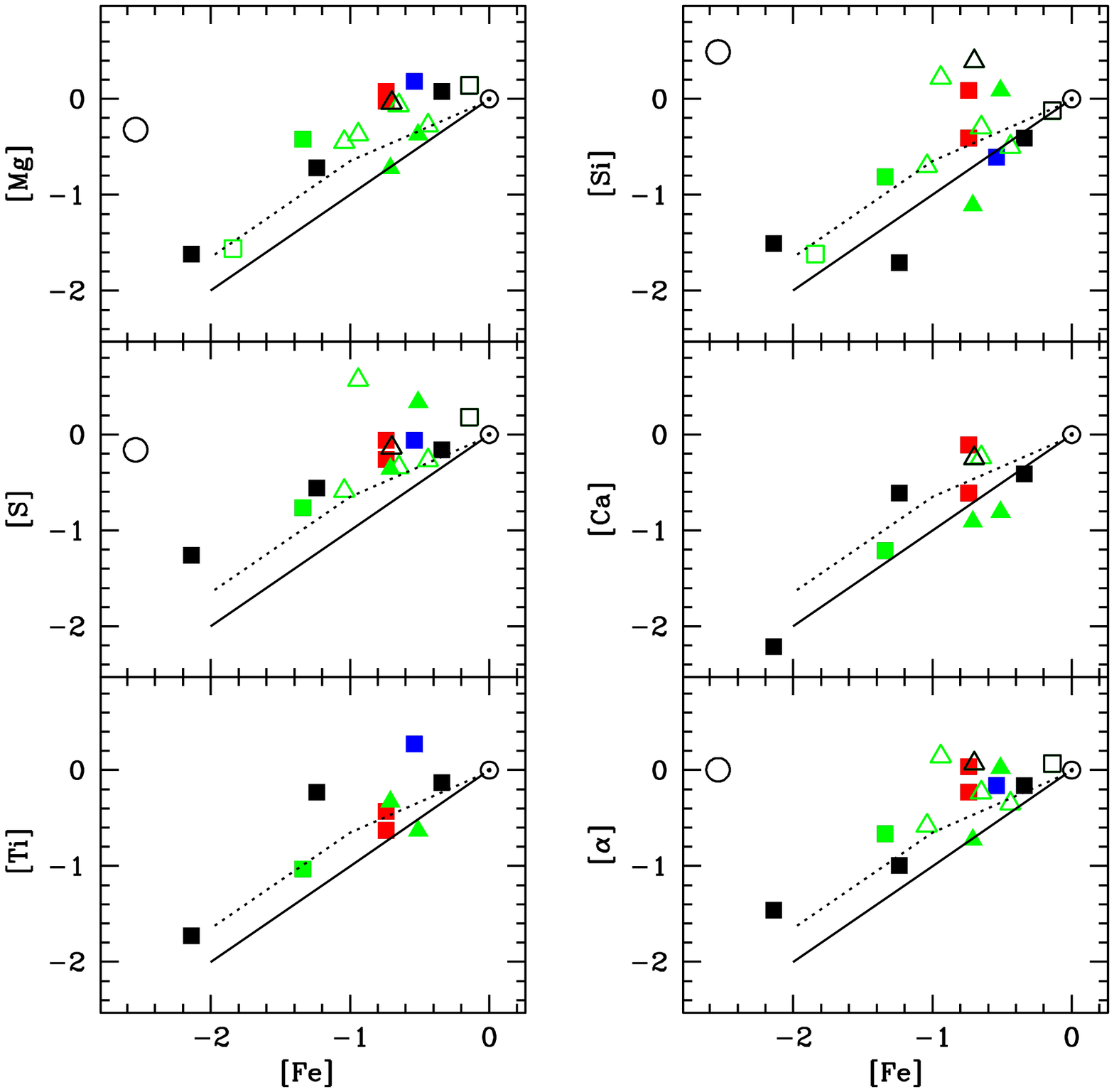}
\caption{[Mg], [Si], [S], [Ca], [Ti], and [$\alpha$] vs [Fe]. 
Our sample of seven EHes is represented by filled squares. 
Two cool EHes analysed by \citet{pan01} are represented 
by filled triangles. The results taken from the literature for 
the EHes with C/He of about 1\% and much lower C/He are represented 
by open triangles and open squares, respectively. DY\,Cen is 
represented by open circle. $\odot$ represents Sun. [X] = [Fe] are  
denoted by the solid lines where X represents Mg, Si, S, Ca, Ti, 
and $\alpha$. The dotted lines are from the relation [X/Fe] 
versus [Fe/H] for normal disk and halo stars \citep{ryd04,gos00}.
\label{fig11}}
\end{figure}
\clearpage
Aluminum is another possible representative of initial metallicity. 
The Al abundances of the EHes follow the Fe abundances (Figure 12)
with an apparent offset of about 0.4 dex in the Fe abundance. Again,
DY\,Cen is a striking exception, but the other 
minority RCBs have an
Al abundance in line with the general Al -- Fe trend for the RCBs
\citep{asp00}. Note that, minority RCBs show lower Fe abundance and 
higher Si/Fe and S/Fe ratios than majority RCBs \citep{lamb94}. 
\citet{pan01} found higher Si/Fe and S/Fe ratios for the Fe-poor 
cool EHe FQ\,Aqr than majority RCBs. But, from our adopted abundances (Table 4)
for FQ\,Aqr, the Si/Fe and S/Fe ratios for FQ\,Aqr and majority RCBs
are in concert. 
\clearpage

\begin{figure}
\epsscale{1.00}
\plotone{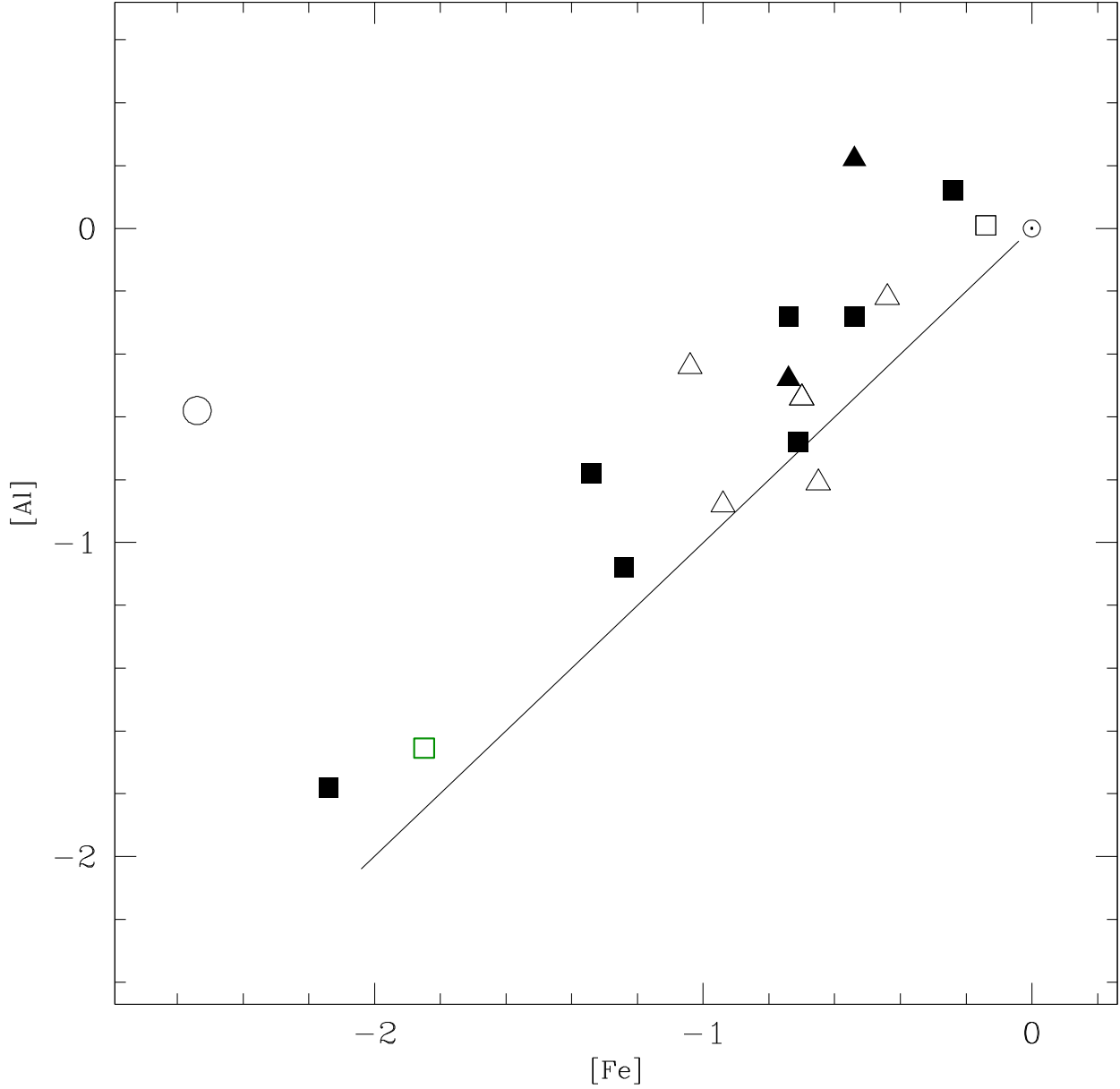}
\caption{[Al] vs [Fe].
Our sample of seven EHes is represented by filled squares.
Two cool EHes analysed by \citet{pan01} are represented
by filled triangles. The results taken from the literature for
the EHes with C/He of about 1\% and much lower C/He are represented
by open triangles and open squares, respectively. DY\,Cen is
represented by open circle. $\odot$ represents Sun. [Al] = [Fe] is
denoted by the solid line.
\label{fig12}}
\end{figure}
\clearpage

In summary, several elements appear to be representative of initial
metallicity. We take Fe for spectroscopic convenience as the representative
of initial metallicity for the EHes but note the dramatic case of DY\,Cen.
The representative of initial metallicity is used to predict 
the initial abundances of elements affected by nuclear reactions and mixing.
\citet{pan01} used Si and S as the representative of initial metallicity 
to derive the initial metallicity M$\equiv$Fe for the EHes. The initial 
metallicity M rederived from an EHe's adopted Si and S abundances is consistent 
with its adopted Fe abundance.

\subsection{Elements affected by evolution}

{\it Hydrogen} --  Deficiency
of H shows a great range over the extended sample of
EHes. The three least H-deficient stars are DY\,Cen, the hot RCB, and
HD\,144941 and V652\,Her,
the two EHes with a very low C abundance (see next section). The remaining
EHe stars have H abundances $\log\epsilon$(H) in the range 5 to 8.
There is a suggestion of a trend of increasing H with increasing
$T_{\rm eff}$ but the hottest EHe LS IV$+6^\circ 2$ does not fit the trend.

{\it Carbon} -- The  
carbon abundances of our septet span a small but definite range:
the mean C/He ratio is 0.0074 with a range from
C/He = 0.0029 for FQ\,Aqr to 0.014 for V1920\,Cyg. The mean C/He from
eight of the ten additional EHes including DY\,Cen is 0.0058 with
a range from 0.0029 to 0.0098. The grand mean from 15 stars is
C/He = 0.0066. 
Two EHes -- HD\,144941 and V652\,Her -- have  much lower C/He
ratios: C/He $= 1.8 \times 10^{-5}$ and $4.0 \times 10^{-5}$ for
HD\,144941 \citep{har97} and V652\,Her \citep{jeff99}, respectively.
This difference in the C/He ratios for EHes between the majority
with C/He of about 0.7 per cent and HD\,144941 and V652\,Her
suggests that a minimum of two mechanisms create EHes.

{\it Nitrogen} -- Nitrogen
is clearly enriched in the great majority of EHes
above its initial abundance expected according to the Fe abundance.
Figure 13 (left-hand panel)  shows that the N abundance for all but 3 of the
17 stars follows the trend expected by the almost complete conversion of the
initial C, N, and O to N through extensive running of the
H-burning CN-cycle and the ON-cycles. The exceptions are
again DY\,Cen (very N-rich for its Fe abundance) and HD\,144941,
one of two stars with a very low C/He ratio,
and LSS\,99, both with a N abundance indicating little N
enrichment over the star's initial N abundance.

{\it Oxygen} -- Oxygen
abundances relative to Fe
range from underabundant by more than 1 dex to overabundant
by almost 2 dex. The stars fall into two groups. Six stars with
[O] $\geq 0$ stand apart from the remainder of the sample for which
the majority (9 of 11) have an
O abundance close to their initial value (Figure  13 (right-hand panel)).
The O/N ratio for this
majority is approximately constant at  O/N $\simeq 1$
and independent of Fe.
 The O-rich stars in order of decreasing Fe abundance are:
LSS\,4357, LSE\,78, V1920\,Cyg, LS IV $-1^\circ2$, FQ\,Aqr, and DY\,Cen.
The very O-poor star (relative to Fe) is V652\,Her, one of two stars with a very
low C/He. The other such star, HD\,144941, has an O (and possibly N)
abundance equal to its
initial value.

A problem is presented by the stars with their O
abundances close to the inferred initial abundances. Eight of the 10 have
an N abundance indicating total conversion of initial C, N, and O
to N via the CNO-cycles, yet the observed O abundance is close to the
initial abundance (unlikely to be just a coincidence but the possibility
needs to be explored).
\clearpage
\begin{figure}
\epsscale{1.00}
\plotone{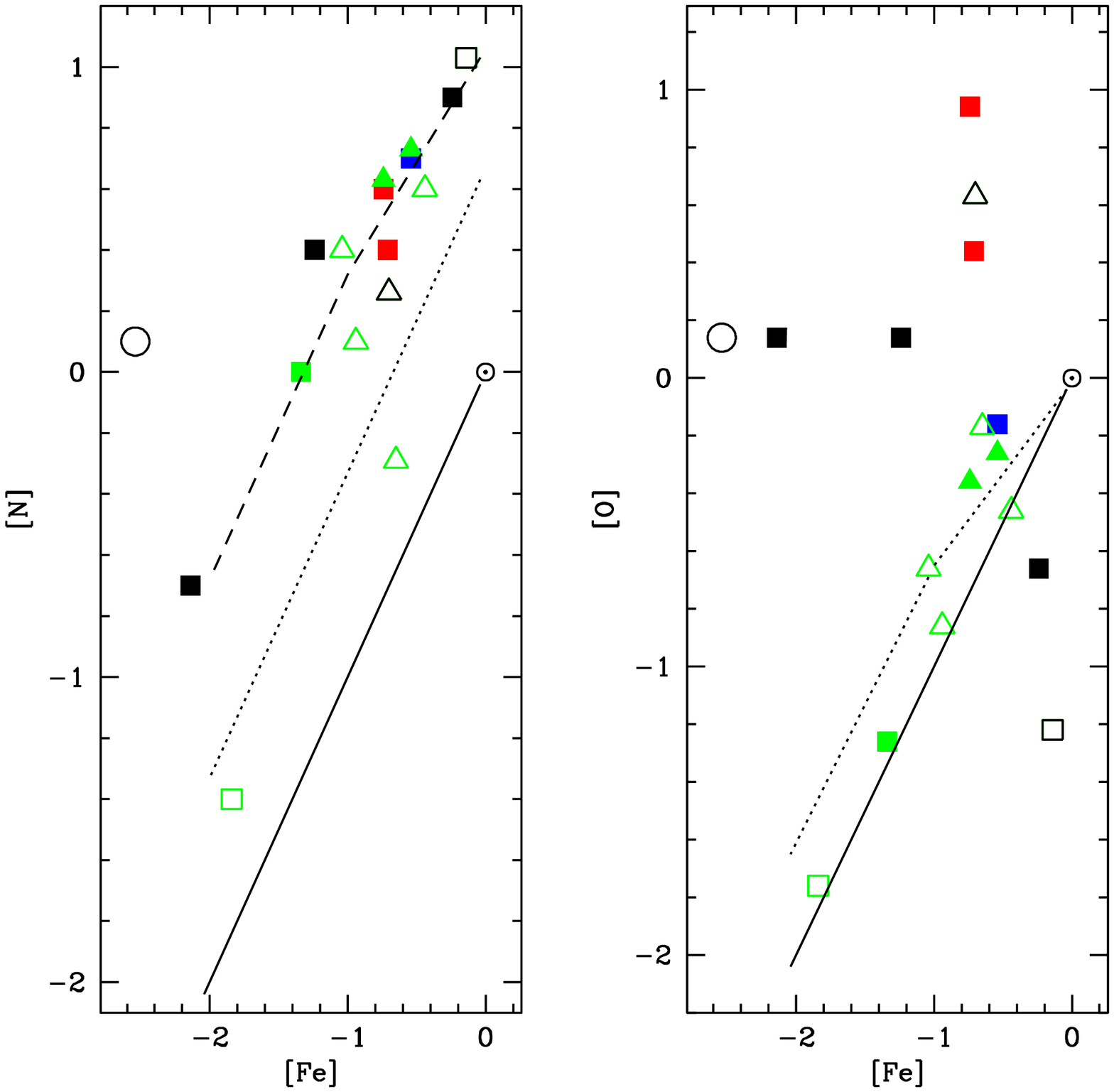}
\caption{Left-hand panel, [N] vs [Fe]. 
Our sample of seven EHes is represented by filled squares.
Two cool EHes analysed by \citet{pan01} are represented
by filled triangles. The results taken from the literature for
the EHes with C/He of about 1\% and much lower C/He are represented
by open triangles and open squares, respectively. DY\,Cen is
represented by open circle. $\odot$ represents Sun. [N] = [Fe] is
denoted by the solid line. The dotted line represents conversion
of the initial sum of C and N to N. The dashed line represents
the locus of the sum of initial C, N, and O converted to N.
Right-hand panel, [O] vs [Fe]. The symbols have the same meaning as in
left-hand panel. [O] = [Fe] is denoted by the solid line. The dotted line is 
from the relation [O/Fe] versus [Fe/H] for normal disk and halo
stars \citep{niss02}.
\label{fig13}}
\end{figure}
\clearpage
{\it Heavy elements} -- Yttrium and Zr abundances
were measured from our {\it STIS} spectra. In
addition, Y and Zr were measured  in the cool EHe LS IV $-14^\circ109$
\citep{pan01}.
Yttrium and Zr abundances are shown in Figure 14 where we
assume that [Zr] = [Fe] represents the initial abundances.
Two stars are severely enriched in Y and Zr: V1920\,Cyg and LSE\,78 with
overabundances of about a factor of 50 (1.7 dex) (see Figure 1 of Pandey et al. 2004). 
Also see Figure 15: the Zr\,{\sc iii} line strength relative to the
Fe\,{\sc ii} line strength is enhanced in Zr enriched stars: LSE\,78 and PV\,Tel,
than the other two stars: FQ\,Aqr and BD+10$^{\circ}$\,2179 with Zr close to
their initial abundance.
This obvious difference in line strengths is also seen in
Figures 1 and 2.
A third star PV\,Tel is enriched by a factor of about 10
(1.0 dex). The other five stars are considered to have their initial
abundances of Y and Zr.
We attribute the occurrence of Y and Zr overabundances to contamination
of the atmosphere by $s$-process products.
\clearpage
\begin{figure}
\epsscale{1.00}
\plotone{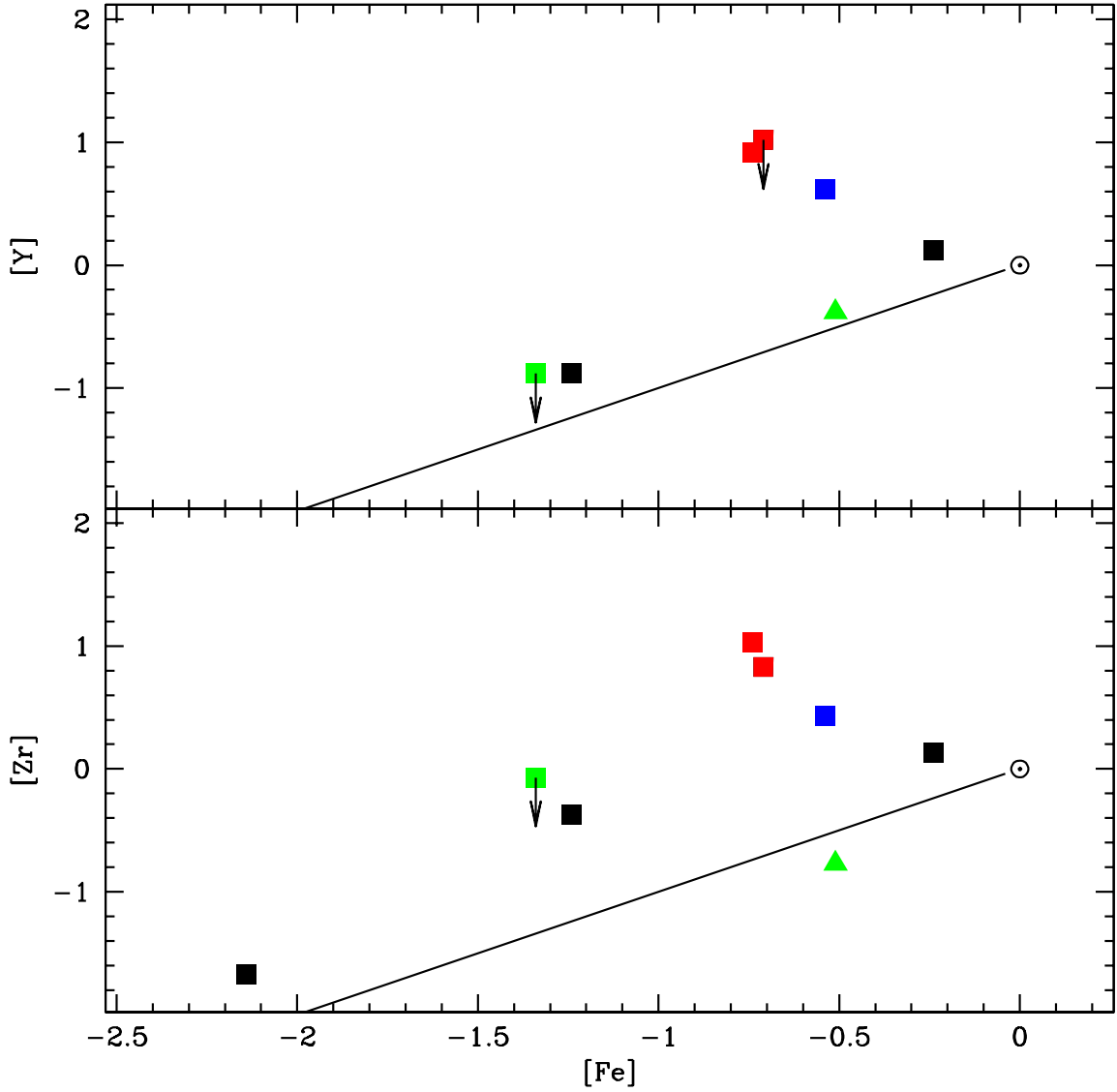}
\caption{[Y] and [Zr] vs [Fe].
Our sample of seven EHes is represented by filled squares.
One of the cool EHes LS IV $-14^\circ109$ analysed by 
\citet{pan01} is represented
by filled triangle. $\odot$ represents Sun. [X] = [Fe] are
denoted by the solid lines where X represents Y and Zr.
\label{fig14}}
\end{figure}

\begin{figure}
\epsscale{1.00}
\plotone{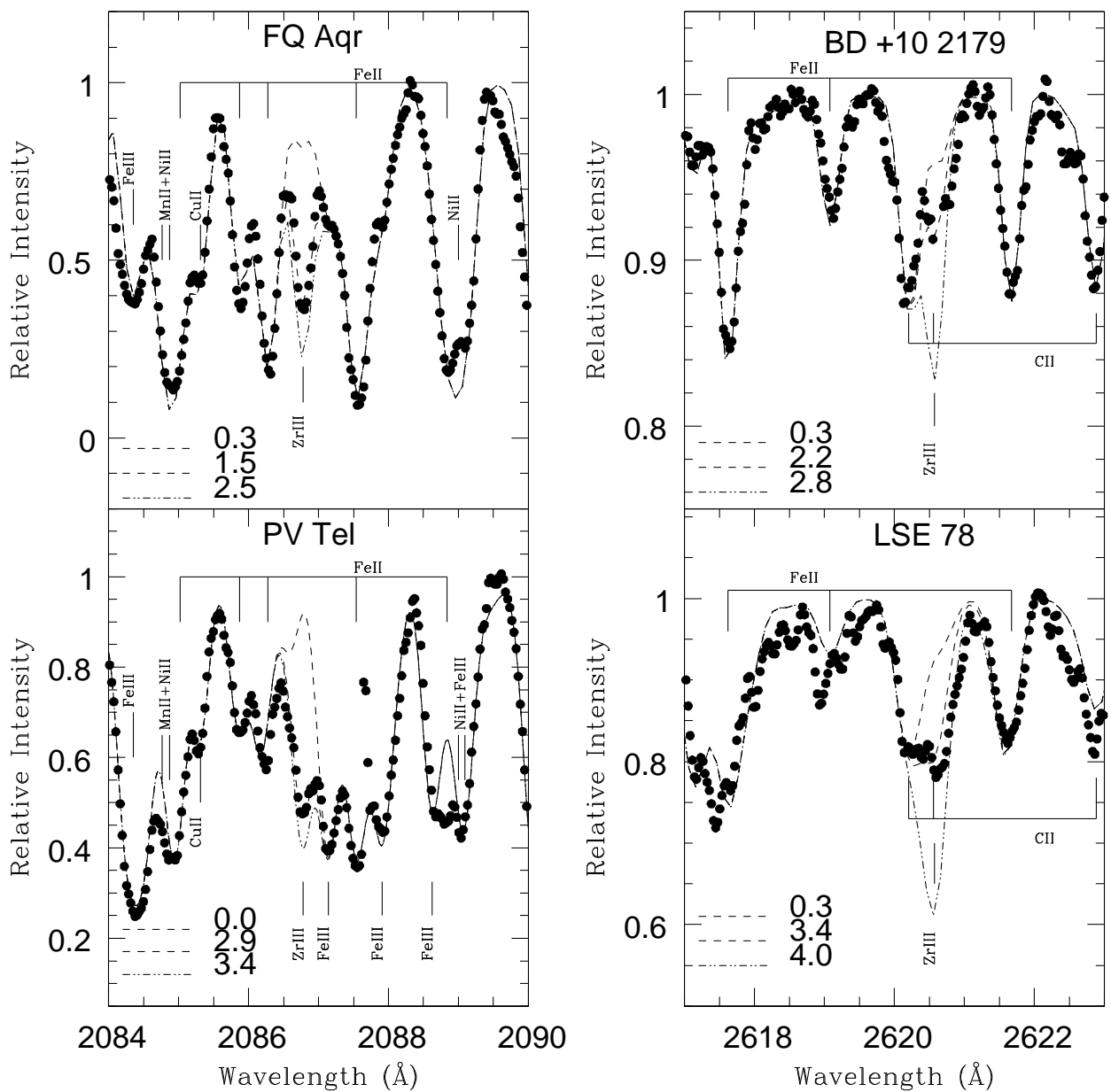}
\caption{The observed spectra of
FQ\,Aqr, PV\,Tel, BD+10$^{\circ}$\,2179, and LSE\,78 are represented by filled
circles. The left-hand panels show the region including the  Zr\,{\sc iii} line
at 2086.78\AA\ for FQ, Aqr and PV\,Tel. The right-hand panels show the region including
the  Zr\,{\sc iii} line at 2620.57\AA\ for BD+10$^{\circ}$\,2179 and LSE\,78.
Synthetic spectra for three different Zr abundances are shown
in each panel for these stars -- see keys on the figure.
In each panel, the principal lines are identified. \label{fig15}}
\end{figure}
\clearpage
The {\it STIS} spectra provide only upper limits for rare-earths
La, Ce, and Nd. In the case of V1920\,Cyg, the Ce and Nd upper limits
suggest an overabundance less than that of Y and Zr, again
assuming that the initial abundances scale directly with the
Fe abundance. For LSE\,78, the La and Ce limits are consistent
with the Y and Zr overabundances. A similar consistency is found
for the Ce abundance in PV\,Tel. The cool EHe LS IV $-14^\circ109$
has a Ba abundance consistent with its initial abundances of Sr,
Y, Zr, and Ba.

\subsection{The R Coronae Borealis stars}

Unlike the EHes where He and C abundances are determined
spectroscopically, the He abundance of the RCBs, except for the rare hot
RCBs, is not measurable. In addition, \citet{asp00} identified that the observed 
strength of a C\,{\sc i} line in RCB's spectrum is considerably lower than the 
predicted and dubbed this `the carbon problem'.  
These factors introduce an uncertainty
into the absolute abundances but Asplund et al. argue that the
abundance ratios, say O/Fe, should be little affected.

The compositions of the RCBs \citep{asp00} show
some similarities to those of the EHes but with differences.
One difference is that
the RCB and EHe metallicity distribution functions are offset by about
0.5 dex in Fe:  the most Fe-rich RCBs have an Fe abundance
about 0.5 dex less than their EHe counterparts.
These offsets differ from element to element: e.g., the Ni distributions 
are very similar but the Ca distributions are offset similarly to Fe. These odd
differences may be reflections of the inability to understand and resolve
the carbon problem.

Despite these differences, there are similarities that support the
reasonable view that the EHe and RCB stars are closely related.
For example, RCBs' O abundances fall into the two groups identified from
a set of O-rich stars and a larger group with O close to the
initial abundance. Also, a few RCBs are $s$-process enriched. Minority
RCBs resemble DY\,Cen, which might be regarded first as RCB rather than an
EHe. It is worthy of note that a few RCBs are known to be
rich in lithium, which must be of recent manufacture. Lithium is not
spectroscopically detectable in the EHes. In this context 
the search of light elements (Be and B) in the $STIS$ spectra of EHes
was unsuccessful. B\,{\sc iii} lines at 2065.776\AA, and at 2067.233\AA\ 
are not detected in EHes' $STIS$ spectra. However, B\,{\sc iii} line at 2065.776\AA\
gives an upper limit to the Boron abundance of about 0.6 dex
for BD $+10^\circ$ 2179. B\,{\sc iii} line at 2067.233\AA\ is severely blended 
by Fe\,{\sc iii} line.


\section{Merger of a He and a CO white dwarf}

The expected composition of a EHe star resulting from the accretion of a helium
white dwarf by a carbon-oxygen white dwarf  was discussed
by Saio \& Jeffery (2002). This scenario is a leading explanation for
EHes and RCBs for reasons of chemical composition and other fits to
observations \citep{asp00,pan01,saio02}.
Here, we examine afresh the  evidence from the
EHes' compositions supporting the merger hypothesis.

In what follows, we consider the initial conditions and the
mixing recipe adopted by Saio \& Jeffery
(2002; see also Pandey et al. 2001).
 The atmosphere and envelope of the resultant EHe
is composed of two zones from the  accreted He white dwarf, and three
zones from the CO white dwarf which is largely undisturbed by the merger.
 Thermal
flashes occur during the accretion phase but the attendant nucleosynthesis
is ignored. We compare the recipe's ability to account for
the observed abundances of H, He, C, N, and O and their run with
Fe. Also, we comment on the $s$-process enrichments.

The He white dwarf contributes  its thin surface layer with
a composition assumed to be the original mix of elements: this layer is
denoted by the label He:H, as in $\beta$(H)$_{\rm{He:H}}$ which is the
mass fraction of hydrogen in the layer of mass $m$(He:H) (in $M_\odot$).
More importantly, the He white dwarf also contributes its He-rich
interior (denoted by the label He:He). Saio \& Jeffery took the
composition of He:He to be CNO-processed, i.e., $\beta$(H) = 0,
$\beta$(He) $\approx$ 1, $\beta$(C) = $\beta$(O) = 0, with  $\beta$(N)
equal to the sum of the initial mass fractions of C, N, and O, and
all other elements at their initial mass fractions.

The CO white dwarf that accretes its companion contributes three
parts to the five part mix. First, a surface layer (denoted by CO:H) with the
original mix of elements. Second, the former He-shell (denoted by CO:He)
with a composition either put identical to that of the He:He layer or
enriched in C and O at the expense of He (see below
for remarks on the layer's $s$-process enrichment). To conclude the
list of ingredients, material from the core may be added
(denoted by CO:CO) with a composition dominated by C and O.

In the representative examples chosen by Saio \& Jeffery (their Table 3), a
0.3$M_\odot$ He white dwarf is accreted by a 0.6$M_\odot$ CO white dwarf
with the accreted material undergoing little mixing with the accretor. The
dominant contributor by mass to the final mix for the envelope
 is the He:He layer with a
mass of 0.3$M_\odot$ followed by the CO:He layer with a mass of
about 0.03$M_\odot$ and the CO:CO layer with a mass of 0.007$M_\odot$
or less. Finally, the surface layers He:H and CO:H with a contribution each of
0.00002$M_\odot$ provide the final mix with a H
deficiency of about 10$^{-4}$.

The stars -- HD\,144941 and V652\,Her -- with the very low C/He ratio
are plausibly identified as resulting from the merger of a He white
dwarf with a more massive He white dwarf \citep{saio00} and
are not further discussed in detail.

{\it Hydrogen --} Surviving hydrogen is contributed by the layers
He:H and CO:H. The formal expression for the mass fraction of H,
Z(H), in the EHe atmosphere is
${\rm Z(H)} = (\beta{\rm (H)}m_{\rm He:H} + \beta{\rm (H)}m_{\rm CO:H})/ M_{tot}$
where $M_{tot}$ is the total mass of  the five contributing layers, and
$\beta$(H)$_{\rm He:H}$ and $\beta$(H)$_{\rm CO:H}$
are expected to be similar and equal to about 0.71. Thus, the residual
H abundance of a EHe is  -- obviously -- mainly set by the
ratio of the combined mass of the two H-containing surface layers to the total
mass of the final envelope and atmosphere. It is not difficult to imagine
that these layers can be of low total mass and, hence, that a EHe
may be very H-deficient.

{\it Helium and Carbon --} For the adopted parameters, primarily
$M$(He:He)/$M$(CO:He) $\approx 10$ and $\beta$(He$_{\rm He:He}$) $\simeq
\beta$(He$_{\rm CO:He}$) $\simeq 1$, the helium from the He:He layer effectively
determines the final He abundance. The carbon ($^{12}$C) is provided
either by C from the top of the CO white dwarf (Saio \& Jeffery's
recipe (1) in their Table 3) or from carbon in the CO:He layer as a result of
He-burning (Saio \& Jeffery's recipe (2) in their Table 3).
 It is of
interest to see if the fact that C/He ratio are generally similar
across the EHe sample offers a clue to the source of the carbon.

In recipe (1), the  C/He mass fraction is given approximately by the
ratio \\ $M$(CO:CO)/$M$(He:He) assuming $\beta$(He)$_{\rm He:He} \simeq
\beta$(C)$_{\rm CO:CO} \simeq 1$. Mass estimates of $M_{\rm CO:CO} \simeq
0.007$ and $M_{\rm He:He} \simeq 0.3$ (Saio \& Jeffery 2002) give
the number ratio C/He $\simeq 0.008$, a value close to the mean of the EHe
sample. 

In recipe (2) where the synthesised C is in the CO:He shell and the contribution
by mass of the CO:CO layers is taken as negligible, the  C/He mass
fraction is approximately $\beta$(C)$_{\rm CO:He}/\beta$(He)$_{\rm He:He}
\times M$(CO:He)/$M$(He:He). Again (of course), substitution from
Saio \& Jeffery's Table 3 gives a number ratio for C/He that is
at the mean observed value.

{\it Nitrogen --} The nitrogen ($^{14}$N) is provided by the He:He and
CO:He layers, principally the former on account of its ten times
greater contribution to the total mass. Ignoring the CO:He layer,
the N mass fraction is given by
$Z$(N) = $\beta$(N)$_{\rm He:He}M_{\rm He:He}/M_{tot}$ and the
mass ratio N/He is given very simply as
$Z$(N)/$Z$(He) = $\beta$(N)$_{\rm He:He}/\beta$(He)$_{\rm He:He}$.
Not only is this ratio independent of the contributions of the
various layers (within limits) but it is directly calculable from the initial
abundances of C, N, and O which depend on the initial Fe abundance.
This prediction which closely matches the observed N and He abundances at all
Fe for all but three stars requires almost complete conversion of initial
C, N, and O to N, as assumed for the layer He:He.

{\it Oxygen --} The oxygen ($^{16}$O)
 is assumed to be a product of He-burning and
to be contributed by either the CO:CO layer (recipe 1) or the CO:He
layer (recipe 2).
Since C and O are contributed by the same layer
 in both recipes, the O/C ratio is
set by a simple ratio of mass fractions: $Z$(O)/$Z$(C) = $\beta$(O)$_{\rm CO:CO}
/\beta$(C)$_{\rm CO:CO}$ for recipe 1, and 
$\beta$(O)$_{\rm CO:He}$/$\beta$(C)$_{\rm CO:He}$ for recipe 2. 
Saio \& Jeffery adopt the ratio $\beta$(O)/$\beta$(C) = 0.25 for both layers
 from models of AGB stars, and, hence, one obtains
the predicted O/C $= 10^{-0.7}$, by number. This is probably insensitive
to the initial metallicity of the AGB star.

The observed O/C across the sample of 15 EHes has a central
value close to the prediction. Extreme values range from O/C =
$10^{0.9}$ for V652\,Her (most probably not the result of a He-CO
merger), also possessing unusually low O, to $10^{-1.9}$ for BD $+10^\circ$ 2179.
If these odd cases are dropped, the mean for
the other 15 is O/C $= 10^{-0.5}$, a value effectively the predicted
one.
The spread from $10^{+0.2}$ to $10^{-1.3}$ corresponding to a large range in
the ratio of the O and C mass fractions from the contributing layer
 exceeds the assessed errors of
measurement.  The spread in O/C is dominated by that in O.
 For the group of  six most oxygen rich EHes, the
observed O/C ratios imply a ratio of the $\beta$s of slightly less than
unity.
The O abundance
for most of the other EHes appears to be a star's initial
abundance. Although one may design a ratio of the $\beta$s that is
metallicity dependent to account for this result, it is then odd that the
O abundances follow the initial O -- Fe relation.

This oddity is removed {\it if} the observed O abundances are indeed
the initial values. This, of course, implies that O is preserved in the
He:He layer,  but, in considering nitrogen, we  noted that
the observed N abundances followed the trend corresponding to
conversion of initial C, N, and O to N in the He:He layer.
Since the ON-cycles operate at a higher temperature than the CN-cycle,
conversion of C to N but not O to N is possible at `low'
temperatures. Additionally at low temperatures and low metallicity, the
$pp$-chain may convert all H to He  before the slower running ON-cycles
have reduced the O abundance to its equilibrium value.
If this speculation is to fit the observations, we must suppose that
the measured N abundances are overestimated by about 0.3 dex in order that
the N abundances be close to the sum of the initial C and N abundances.
It remains to be shown that the  He:He layer of a
He white dwarf can be created by H-burning by the $pp$-chains
and the CN-cycle and without operation of the ON-cycle.

Were the entire He:He layer exposed to the temperatures for ON-cycling,
the reservoir of $^3$He needed to account for Li in some
RCBs would be destroyed. The $^3$He is a product of main sequence
evolution where the $pp$-chain partially operates well outside the
H-burning core. This $^3$He is then later converted to $^7$Li
by the \citet{cam71} mechanism: $^3$He($^4$He,$\gamma)^7$Be$(e^-,\nu)^7$Li. 
The level of the Li abundance, when present, is such that
large-scale preservation of $^3$He seems necessary prior to the onset
of the Cameron-Fowler mechanism. This is an indirect indication that the
He:He layer was not in every case heated such that the CNO-cycles
converted all C,N, and O to N. (Lithium production through spallation 
reactions on the stellar surface is not an appealing alternative. One
unattractive of spallation is that it results in a ratio
$^7$Li/$^6$Li $\sim 1$ but observations suggest that the
observed lithium is almost pure $^7$Li.)

{\it Yttrium and Zirconium --} The
$s$-process enrichment is sited in the CO:He and CO:CO layers.
Saio \& Jeffery assumed an enrichment by a factor of 10 in the
CO:He.
This factor and
the  small mass ratio $M$(CO:He)$/M_{tot}$ result in very
little enrichment for the EHe.
Observed Y and Zr enrichments require either a greater enrichment in
the CO:He layer or addition of material from the CO:CO layer. Significantly,
the two most obviously  $s$-process enriched EHes are also among the
most O-rich.

\section{Concluding remarks}

This LTE model atmosphere
analysis of high-resolution {\it STIS} spectra undertaken primarily
to investigate the abundances of $s$-process elements in the
EHe stars has shown that indeed a few EHes exhibit marked overabundances
of Y and Zr.  The {\it STIS} spectra additionally provide abundances
of other elements and, in particular, of several Fe-group elements
not observable in optical spectra. We combine the results of the
{\it STIS} analysis with  abundance analyses
based on newly obtained or published  optical spectra. Our results
for seven EHes and approximately  24 elements per star
are supplemented with abundances taken from the literature for
an additional ten EHes. The combined sample of 17 stars with
abundances obtained in a nearly uniform manner provides the most
complete dataset yet obtained for these very H-deficient
stars.

Our interpretation of the EHe's atmospheric compositions
considers simple recipes based  on the idea that the
EHe is a consequence of the accretion of a He white
dwarf by a more massive CO white dwarf. (Two stars of
low C/He ratio are more probably a result of the merger of two
He white dwarfs.) These recipes
adapted from Saio \& Jeffery (2002) are quite successful.
A EHe's initial composition is inferred from the measured
Fe abundance, but other elements from Al to Ni could equally
well be identified as the representative of initial metallicity. 
Saio \& Jeffery's
recipes plausibly account for the H, He, C, and N abundances
and for the O abundance of a few stars. Other stars show an
O abundance similar to the expected initial abundance. This
similarity would seem to require that the He-rich material of
the He white dwarf was exposed to the CN-cycle but not the
ON-cycles.

Further progress in elucidating the origins of the EHes from
determinations of their chemical compositions 
requires two principal developments. First, the
abundance analyses should be based on Non-LTE atmospheres and
Non-LTE line formation. The tools to implement these two
steps are available but limitations in available atomic data
may need to be addressed. In parallel with this work, a continued
effort should be made to include additional elements. Neon is
of particular interest as $^{22}$Ne is produced from  $^{14}$N
by $\alpha$-captures
prior to the onset of He-burning by the $3\alpha$-process.
Hints of Ne enrichment exist \citep{pan01}. Second,
a rigorous theoretical treatment of the merger of the He
white dwarf with the CO white dwarf must be developed
with inclusion of the hydrodynamics and the nucleosynthesis
occurring during and following the short-lived accretion
process. A solid beginning has been made in this
direction, see, for example \citet{gue04}.

There remains the puzzling case of DY\,Cen and the minority
RCBs \citep{lamb94} with their highly anomalous
composition. Are these anomalies the result of very
peculiar set of nuclear processes? Or has the `normal' composition
of a RCB been altered by fractionation in the atmosphere or
circumstellar shell?

\acknowledgments

This research was supported by the Robert A. Welch Foundation, Texas,
and the Space Telescope Science Institute grant GO-09417. 
GP thanks Simon Jeffery for the travel support and the hospitality at 
Armagh Observatory where a part of this work was carried out. GP also
thanks Baba Verghese for the help provided in installing the LTE code,
and the referee Uli Heber for his encouraging remarks.

\appendix
\section{Appendix material: Lines used for abundance analysis}

\noindent The lines used for the abundance analysis of LSE\,78, BD\,+10$^{\circ}$\,2179,
V1920\,Cyg, HD\,124448, PV\,Tel, LS\,IV-1$^{\circ}$\,2, and FQ\,Aqr are
given in the Tables 11 to 20 (available only in electronic edition). The $gf$-value, 
the lower excitation potential ($\chi$), log of Stark damping constant/electron number
density ($\Gamma_{el}$), log of radiative damping constant ($\Gamma_{rad}$), 
and the abundance (log $\epsilon$) derived
for each line are listed. Also listed are the equivalent widths ($W_{\lambda}$)
corresponding to the abundances derived by spectrum synthesis for most individual lines. 
The symbol ? follows uncertain abundances, and `Synth' in the 
sixth column implies spectrum synthesis.

{\bf References}

\noindent Artru = Artru et al. (1981) \\
\noindent CL = Crespo Lopez-Urrutia et al. (1994) \\
\noindent Ekberg = Ekberg (1997) \\
\noindent Jeffery = Jeffery, Woolf, \& Pollacco (2001) \\
\noindent Kurucz = Kurucz's database \\
\noindent Luck = Compilations by R. E. Luck \\
\noindent NIST = NIST database \\
\noindent Pandey = Pandey et al. (2004) \\
\noindent RU = Raassen \& Uylings (1997) \\
\noindent Salih = Salih, Lawler, \& Whaling (1985) \\
\noindent UR = Uylings \& Raassen (1997) \\
\noindent WFD = Wiese, Fuhr, \& Deters (1996) \\

\clearpage







%
%
%

\begin{deluxetable}{lccccccc}
\tabletypesize{\small}
\tablewidth{0pt}
\tablecolumns{8}
\setcounter{table} {10}
\tablecaption{Ultraviolet lines used to derive elemental abundances for LSE\,78
with the model atmosphere (18300, 2.2, 16.0)}
\tablehead{
\multicolumn{1}{l}{Ion} & \colhead{} & \colhead{} & \colhead{} & \colhead{} \\
\multicolumn{1}{l}{$\lambda$(\AA)} & \multicolumn{1}{c}{log $gf$} & \multicolumn{1}{c}{$\chi$(eV)} &
\multicolumn{1}{c}{$\Gamma_{el}$} & \multicolumn{1}{c}{$\Gamma_{rad}$} & 
\multicolumn{1}{c}{$W_{\lambda}$(m\AA)} &
\multicolumn{1}{c}{log $\epsilon$} & \multicolumn{1}{c}{Ref $^a$} \\
}
\startdata
He\,{\sc i} & & & & & &&\\
2652.566 &--1.410 &3.267 &  &  & Synth &    11.54 & Kurucz  \\
C\,{\sc ii} & & & & & &&\\
1988.530 &--2.123 & 16.333 &  &  &108&     9.4  & WFD  \\
2092.165 &--1.473 & 20.704 &  &  &58&     9.6  & WFD  \\
2173.849 &--1.185 & 14.449 &--4.930 &  9.050 &Synth&     9.3? & WFD  \\
2174.168 &--1.486 & 14.449 &--4.930 &  9.050 &Synth&     9.3? & WFD  \\
2269.688 &--0.810 & 22.537 &  &  &86&     9.8  & WFD  \\
\enddata
\tablenotetext{a}{Sources of $gf$-values.}
\end{deluxetable}

\begin{deluxetable}{lccccccc}
\tabletypesize{\small}
\tablewidth{0pt}
\tablecolumns{8}
\tablecaption{Ultraviolet lines used to derive elemental abundances for BD\,+10$^{\circ}$\,2179
with the model atmosphere (16900, 2.55, 4.5)}
\tablehead{
\multicolumn{1}{l}{Ion} & \colhead{} & \colhead{} & \colhead{} & \colhead{} \\
\multicolumn{1}{l}{$\lambda$(\AA)} & \multicolumn{1}{c}{log $gf$} & \multicolumn{1}{c}{$\chi$(eV)} &
\multicolumn{1}{c}{$\Gamma_{el}$} & \multicolumn{1}{c}{$\Gamma_{rad}$} &
\multicolumn{1}{c}{$W_{\lambda}$(m\AA)} &
\multicolumn{1}{c}{log $\epsilon$} & \multicolumn{1}{c}{Ref $^a$} \\
}
\startdata
C\,{\sc ii} & & & &&&&\\
1916.010 &--2.369 & 14.449 &  &  &59&     8.9   & WFD \\
1926.770 &--2.366 & 13.716 &  &  &Synth&     9.5   & WFD \\
1927.020 &--2.065 & 13.716 &  &  &Synth&     9.5   & WFD \\
1988.530 &--2.123 & 16.333 &  &  &78&     9.6   & WFD \\
2017.929 &--1.793 & 16.332 &  &  8.600 &Synth&     9.4   & WFD \\
2018.379 &--1.492 & 16.333 &  &  8.600 &Synth&     9.3?  & WFD \\
2092.165 &--1.473 & 20.704 &  &  &39&     9.6   & WFD \\
\enddata
\tablenotetext{a}{Sources of $gf$-values.}
\end{deluxetable}

\begin{deluxetable}{lccccccc}
\tabletypesize{\small}
\tablewidth{0pt}
\tablecolumns{8}
\tablecaption{Optical lines used to derive elemental abundances for BD\,+10$^{\circ}$\,2179
with the model atmosphere (16400, 2.35, 6.5)}
\tablehead{
\multicolumn{1}{l}{Ion} & \colhead{} & \colhead{} & \colhead{} & \colhead{} \\
\multicolumn{1}{l}{$\lambda$(\AA)} & \multicolumn{1}{c}{log $gf$} & \multicolumn{1}{c}{$\chi$(eV)} &
\multicolumn{1}{c}{$\Gamma_{el}$} & \multicolumn{1}{c}{$\Gamma_{rad}$} &
\multicolumn{1}{c}{$W_{\lambda}$(m\AA)} &
\multicolumn{1}{c}{log $\epsilon$} & \multicolumn{1}{c}{Ref $^a$} \\
}
\startdata
H\,{\sc i} & & & &&&&\\
4101.734 &--0.753 & 10.150 &    &8.790 &Synth&     8.2   &Jeffery \\
4340.462 &--0.447 & 10.150 &    &8.790 &Synth&     8.2   &Jeffery \\
4861.323 &--0.020 & 10.150 &    &8.780 &Synth&     8.5   &Jeffery \\
\hline
Mean:    &        &         &          & &&&8.30$\pm$0.17$\pm$0.20\\
\hline
He\,{\sc i} & & & &&&&\\
3819.600 &--2.959 & 20.964 & &   &Synth&  11.54    & Jeffery \\
3819.600 &--1.790 & 20.964 & &   &Synth&  11.54    & Jeffery \\
\enddata
\tablenotetext{a}{Sources of $gf$-values.}
\end{deluxetable}

\begin{deluxetable}{lccccccc}
\tabletypesize{\small}
\tablewidth{0pt}
\tablecolumns{8}
\tablecaption{Ultraviolet lines used to derive elemental abundances for V1920\,Cyg
with the model atmosphere (16300, 1.7, 15.0)}
\tablehead{
\multicolumn{1}{l}{Ion} & \colhead{} & \colhead{} & \colhead{} & \colhead{} \\
\multicolumn{1}{l}{$\lambda$(\AA)} & \multicolumn{1}{c}{log $gf$} & \multicolumn{1}{c}{$\chi$(eV)} &
\multicolumn{1}{c}{$\Gamma_{el}$} & \multicolumn{1}{c}{$\Gamma_{rad}$} &
\multicolumn{1}{c}{$W_{\lambda}$(m\AA)} &
\multicolumn{1}{c}{log $\epsilon$} & \multicolumn{1}{c}{Ref $^a$} \\
}
\startdata
He\,{\sc i} & & & &&&&\\
2652.566 &--1.410 &3.267   &  &  &Synth&    11.5  & Kurucz  \\
C\,{\sc ii} & & & &&&&\\
1988.530 &--2.123 & 16.333 &  &  &Synth&     9.4?   & WFD \\
2092.165 &--1.473 & 20.704 &  &  &Synth&    10.0?   & WFD \\
2137.896 &--0.735 & 16.333 &  &  &Synth&     9.2    & WFD \\
2137.926 &--1.690 & 16.333 &  &  &Synth&     9.2    & WFD \\
2188.375 &--1.014 & 20.922 &  &  &83&     9.6    & WFD \\
\enddata
\tablenotetext{a}{Sources of $gf$-values.}
\end{deluxetable}

\begin{deluxetable}{lccccccc}
\tabletypesize{\small}
\tablewidth{0pt}
\tablecolumns{8}
\tablecaption{Optical lines used to derive elemental abundances for V1920\,Cyg
with the model atmosphere (16330, 1.8, 20.0)}
\tablehead{
\multicolumn{1}{l}{Ion} & \colhead{} & \colhead{} & \colhead{} & \colhead{} \\
\multicolumn{1}{l}{$\lambda$(\AA)} & \multicolumn{1}{c}{log $gf$} & \multicolumn{1}{c}{$\chi$(eV)} &
\multicolumn{1}{c}{$\Gamma_{el}$} & \multicolumn{1}{c}{$\Gamma_{rad}$} &
\multicolumn{1}{c}{$W_{\lambda}$(m\AA)} &
\multicolumn{1}{c}{log $\epsilon$} & \multicolumn{1}{c}{Ref $^a$} \\
}
\startdata
H\,{\sc i} & & & &&&&\\
6562.852 &0.710  & 10.199 &  &  8.760 &$<$8&  $<$6.2   &Jeffery \\
\hline
Mean:    &        &         &          & &&&$<$6.20$\pm$0.00$\pm$0.16\\
\hline
He\,{\sc i} & & & &&&&\\
3819.600 &--2.959 & 20.964 & &   &Synth   &11.50    & Jeffery \\
3819.600 &--1.790 & 20.964 & &   &Synth   &11.50    & Jeffery \\
3819.600 &--1.040 & 20.964 & &   &Synth   &11.50    & Jeffery \\
3819.610 &--1.790 & 20.964 & &   &Synth   &11.50    & Jeffery \\
\enddata
\tablenotetext{a}{Sources of $gf$-values.}
\end{deluxetable}

\begin{deluxetable}{lccccccc}
\tabletypesize{\small}
\tablewidth{0pt}
\tablecolumns{8}
\tablecaption{Ultraviolet lines used to derive elemental abundances for HD\,124448
with the model atmosphere (16100, 2.3, 10.0)}
\tablehead{
\multicolumn{1}{l}{Ion} & \colhead{} & \colhead{} & \colhead{} & \colhead{} \\
\multicolumn{1}{l}{$\lambda$(\AA)} & \multicolumn{1}{c}{log $gf$} & \multicolumn{1}{c}{$\chi$(eV)} &
\multicolumn{1}{c}{$\Gamma_{el}$} & \multicolumn{1}{c}{$\Gamma_{rad}$} &
\multicolumn{1}{c}{$W_{\lambda}$(m\AA)} &
\multicolumn{1}{c}{log $\epsilon$} & \multicolumn{1}{c}{Ref $^a$} \\
}
\startdata
C\,{\sc ii} & & & &&&&\\
2092.165 &--1.473 & 20.704  &  &  &35&    9.42  & WFD \\
2137.896 &--0.735 & 16.333  &  &  &Synth&    9.21  & WFD \\
2137.926 &--1.690 & 16.333  &  &  &Synth&    9.21  & WFD \\
2188.375 &--1.014 & 20.922  &  &  &63&    9.52?  & WFD \\
2269.688 &--0.810 & 22.537  &  &  &26&    9.22  & WFD \\
2574.770 &--0.620 & 18.046  &  &  &Synth&    9.21  & WFD \\
2574.865 &--0.470 & 18.046  &  &  &Synth&    9.21  & WFD \\
\enddata
\tablenotetext{a}{Sources of $gf$-values.}
\end{deluxetable}

\begin{deluxetable}{lccccccc}
\tabletypesize{\small}
\tablewidth{0pt}
\tablecolumns{8}
\tablecaption{Optical lines used to derive elemental abundances for HD\,124448
with the model atmosphere (15500, 1.9, 12.0)}
\tablehead{
\multicolumn{1}{l}{Ion} & \colhead{} & \colhead{} & \colhead{} & \colhead{} \\
\multicolumn{1}{l}{$\lambda$(\AA)} & \multicolumn{1}{c}{log $gf$} & \multicolumn{1}{c}{$\chi$(eV)} &
\multicolumn{1}{c}{$\Gamma_{el}$} & \multicolumn{1}{c}{$\Gamma_{rad}$} &
\multicolumn{1}{c}{$W_{\lambda}$(m\AA)} &
\multicolumn{1}{c}{log $\epsilon$} & \multicolumn{1}{c}{Ref $^a$} \\
}
\startdata
H\,{\sc i} & & & &&&&\\
6562.852 &0.710  & 10.199 &    &8.760 &$<$21&  $<$6.3   &Jeffery \\
\hline
Mean:    &        &         &          & &&&$<$6.30$\pm$0.00$\pm$0.20\\
\hline
He\,{\sc i} & & & &&&&\\
6678.150 &0.330  & 21.218 &--4.520 &  &Synth& 11.53     &Kurucz \\
C\,{\sc ii} & & & &&&&\\
5537.609 &--1.793 & 19.495 &  &  8.940 &45& 8.98      & WFD \\
6578.052 &--0.026 & 14.449 &--4.340  & 9.070 &598& 9.13      & WFD \\
\enddata
\tablenotetext{a}{Sources of $gf$-values.}
\end{deluxetable}

\begin{deluxetable}{lccccccc}
\tabletypesize{\small}
\tablewidth{0pt}
\tablecolumns{8}
\tablecaption{Ultraviolet lines used to derive elemental abundances for PV\,Tel
with the model atmosphere (13750, 1.6, 9.0)}
\tablehead{
\multicolumn{1}{l}{Ion} & \colhead{} & \colhead{} & \colhead{} & \colhead{} \\
\multicolumn{1}{l}{$\lambda$(\AA)} & \multicolumn{1}{c}{log $gf$} & \multicolumn{1}{c}{$\chi$(eV)} &
\multicolumn{1}{c}{$\Gamma_{el}$} & \multicolumn{1}{c}{$\Gamma_{rad}$} &
\multicolumn{1}{c}{$W_{\lambda}$(m\AA)} &
\multicolumn{1}{c}{log $\epsilon$} & \multicolumn{1}{c}{Ref $^a$} \\
}
\startdata
C\,{\sc ii} & & & &&&&\\
2137.896 &--0.735 & 16.333  &  &  &Synth&    9.13  & WFD \\
2137.926 &--1.690 & 16.333  &  &  &Synth&    9.13  & WFD \\
2509.124 &--0.767 & 13.716  &  &  9.680 &348&    9.53  & WFD \\
\hline
Mean:    &        &         &          & &&&9.33$\pm$0.28$\pm$0.04\\
\hline
C\,{\sc iii} & & & &&&&\\
1908.730 &--6.729 &  0.000  &  &  2.640 &184&    9.59  & WFD \\
\hline
Mean:    &        &         &          & &&&9.59$\pm$0.28$\pm$0.25\\
\enddata
\tablenotetext{a}{Sources of $gf$-values.}
\end{deluxetable}

\begin{deluxetable}{lccccccc}
\tabletypesize{\small}
\tablewidth{0pt}
\tablecolumns{8}
\tablecaption{Ultraviolet lines used to derive elemental abundances for  LS\,IV-1$^{\circ}$\,2
with the model atmosphere (12750, 1.75, 10.0)}
\tablehead{
\multicolumn{1}{l}{Ion} & \colhead{} & \colhead{} & \colhead{} & \colhead{} \\
\multicolumn{1}{l}{$\lambda$(\AA)} & \multicolumn{1}{c}{log $gf$} & \multicolumn{1}{c}{$\chi$(eV)} &
\multicolumn{1}{c}{$\Gamma_{el}$} & \multicolumn{1}{c}{$\Gamma_{rad}$} &
\multicolumn{1}{c}{$W_{\lambda}$(m\AA)} &
\multicolumn{1}{c}{log $\epsilon$} & \multicolumn{1}{c}{Ref $^a$} \\
}
\startdata
C\,{\sc ii} & & & &&&&\\
2137.896 &--0.735 & 16.333  &  &  &Synth&    9.3  & WFD \\
2137.926 &--1.690 & 16.333  &  &  &Synth&    9.3  & WFD \\
2509.124 &--0.767 & 13.716  &  &  9.680 &445&    9.7  & WFD \\
\hline
Mean:    &        &         &          & &&&9.50$\pm$0.28$\pm$0.06\\
\hline
Mg\,{\sc ii} & & & &&&&\\
2660.754 &--0.480 &  8.864  &--3.510 &  &Synth&    6.9  & NIST \\
2660.756 &--1.780 &  8.864  &--3.510 &  &Synth&    6.9  & NIST \\
\enddata
\tablenotetext{a}{Sources of $gf$-values.}
\end{deluxetable}

\begin{deluxetable}{lccccccc}
\tabletypesize{\small}
\tablewidth{0pt}
\tablecolumns{8}
\tablecaption{Ultraviolet lines used to derive elemental abundances for  FQ\,Aqr
with the model atmosphere (8750, 0.30, 7.5)}
\tablehead{
\multicolumn{1}{l}{Ion} & \colhead{} & \colhead{} & \colhead{} & \colhead{} \\
\multicolumn{1}{l}{$\lambda$(\AA)} & \multicolumn{1}{c}{log $gf$} & \multicolumn{1}{c}{$\chi$(eV)} &
\multicolumn{1}{c}{$\Gamma_{el}$} & \multicolumn{1}{c}{$\Gamma_{rad}$} &
\multicolumn{1}{c}{$W_{\lambda}$(m\AA)} &
\multicolumn{1}{c}{log $\epsilon$} & \multicolumn{1}{c}{Ref $^a$} \\
}
\startdata
C\,{\sc ii} & & & &&&&\\
2018.379    &    --1.492    &    16.333 &  &  8.600 &Synth& 9.3?   & WFD \\
N\,{\sc ii} & & & &&&&\\
2142.775    &    --6.359    &   0.016   &  &  &Synth& 6.7?   & WFD \\
Mg\,{\sc ii} & & & &&&&\\
2660.754    &    --0.48     &   8.864   &--3.510 &  &Synth& 6.0    & NIST \\  
2660.756    &    --1.78     &   8.864   &--3.510 &  &Synth& 6.0    & NIST \\
2660.818    &    --0.63     &   8.864   &--3.510 &  &Synth& 6.0    & NIST \\
\enddata
\tablenotetext{a}{Sources of $gf$-values.}
\end{deluxetable}


\end{document}